\documentclass[aps,prl,twocolumn,superscriptaddress,longbibliography]{revtex4-2}
\usepackage{breakurl}
\usepackage{xcolor}
\usepackage{sidecap}
\usepackage{amssymb}
\usepackage{hhline}
\usepackage{multirow}
\sidecaptionvpos{figure}{t}
\usepackage{amsmath}
\usepackage{mathrsfs}
\usepackage{graphicx}
\usepackage{esint}
\usepackage{epstopdf}
\usepackage{rotating}
\graphicspath{{pict/}{./}}
\usepackage{bm}%
\usepackage{microtype,bm,bbm,graphicx,booktabs,times}

\newcounter{Fig}

\newcommand\mymapstol{\mathrel{\ooalign{$\leftarrow$\cr%
  \kern1.75ex\raise0.275ex\hbox{\scalebox{1}[0.4]{$\mid$}}\cr}}}

\newcommand\mymapstor{\mathrel{\ooalign{$\rightarrow$\cr%
  \kern-.15ex\raise.275ex\hbox{\scalebox{1}[0.4]{$\mid$}}\cr}}}
  
\usepackage{graphicx,tipa}
\newcommand{\arc}[1]{{%
  \setbox9=\hbox{#1}%
  \ooalign{\resizebox{\wd9}{\height}{\texttoptiebar{\phantom{A}}}\cr#1}}}

\begin{document}

\title{Geometric Phase-Driven Scattering Evolutions}
\author{Pengxiang Wang}
\affiliation{School of Optical and Electronic Information, Huazhong University of Science and Technology, Wuhan, Hubei 430074, P. R. China}
\author{Yuntian Chen}
\email{Email: yuntian@hust.edu.cn}
\affiliation{School of Optical and Electronic Information, Huazhong University of Science and Technology, Wuhan, Hubei 430074, P. R. China}
\affiliation{Wuhan National Laboratory for Optoelectronics, Huazhong University of Science and Technology, Wuhan, Hubei 430074, P. R. China}
\author{Wei Liu}
\email{Email: wei.liu.pku@gmail.com}
\affiliation{College for Advanced Interdisciplinary Studies, National University of Defense Technology, Changsha 410073, P. R. China.}
\affiliation{Nanhu Laser Laboratory and Hunan Provincial Key Laboratory of Novel Nano-Optoelectronic Information Materials and
Devices, National University of Defense Technology, Changsha 410073, P. R. China.}
%

\begin{abstract}
Conventional approaches for scattering manipulations rely on the technique of field expansions into spherical harmonics (electromagnetic multipoles), which nevertheless is non-generic (expansion coefficients depend on the position of the  coordinate system's origin) and more descriptive than predictive. Here we explore this classical topic from a different perspective of controlled excitations and interferences of quasi-normal modes (QNMs) supported by the scattering system. Scattered waves are expanded into not spherical harmonics but radiations of QNMs, among which the relative amplitudes and phases are crucial factors to architect for scattering manipulations. Relying on the electromagnetic reciprocity, we provide full geometric representations based on the Poincaré sphere for those factors, and identify the hidden underlying geometric phases of QNMs that drive the scattering evolutions. Further synchronous exploitations of the incident polarization-dependent geometric phases and excitation amplitudes enable efficient manipulations of both scattering intensities and polarizations. Continuous geometric phase spanning $2\pi$ is directly manifest through scattering variations, even in the rather elementary configuration of an individual particle scattering waves of varying polarizations. We have essentially established a profoundly  all-encompassing framework for the calculations of geometric phase in scattering systems, which will greatly broaden horizons of many disciplines not only in photonics but also in general wave physics where geometric phase is generic and ubiquitous.
\end{abstract}

\maketitle

The seminal topic of electromagnetic scatterings by particles has been the cornerstone for investigations of light-matter interactions and various scattering-related applications~\cite{Bohren1983_book,KUZNETSOV_Science_optically_2016,LIU_2018_Opt.Express_Generalized,KRASNOK_2019_Adv.Opt.Photon.AOP_Anomalies,KIVSHAR_2022_NanoLett._Rise,ALTUG_2022_Nat.Nanotechnol._Advances,FAN_2022_Nat.Photon._Photonics}.
It has recently rapidly merged with other vibrant branches of singular, topological and non-hermitian photonics~\cite{LIU_ArXiv201204919Phys._Topological,MIRI_2019_Science_Exceptionala,KOSHELEV_2019_ScienceBulletin_Metaoptics,KANG_2023_NatRevPhys_Applications}. Despite its rather long history and the aforementioned multidisciplinary advances, the central mathematical and physical technique for the field of particle scatterings remains to be spherical harmonics and  electromagnetic multipolar expansions~\cite{Bohren1983_book,DOICU_light_2006}. Indeed many breakthroughs in this field have been made based on this technique, such as recent introductions of Poincar\'{e}-Hopf theorem~\cite{NEEDHAM__Visuala}, electromagnetic multipolar parity~\cite{Liu2014_ultradirectional} and duality~\cite{JACKSON_1998__Classical} into Mie theory to reveal its intrinsic topological and geometric properties~\cite{LIU_Phys.Rev.Lett._generalized_2017,CHEN_2019__Singularities,CHEN_2019_ArXiv190409910Math-PhPhysicsphysics_Linea,FERNANDEZ-CORBATON_2013_Phys.Rev.Lett._Electromagnetica,YANG_2020_ACSPhotonics_Electromagnetic}).

Nevertheless, the language of spherical harmonics and electromagnetic multipoles is more descriptive than predictive: except for some special scenarios of particles with ideal spherical or cylindrical symmetries, this technique describes the already known fields (either the near or scattered far fields calculated with other numerical methods) rather than predict them.  For example, even for the elementary case of plane waves scattered by a particle of arbitrary shape, knowing the multipolar components of one scattering configuration barely tells anything about the scatterings of another even neighboring configuration with a slightly different incident direction and/or polarization. Moreover, the multipolar expansion coefficients are non-generic and highly dependent on the position of the coordinate system chosen~\cite{Bohren1983_book,DOICU_light_2006,JACKSON_1998__Classical}. Formalisms based on spherical harmonics are usually cumbersome and tend to obscure rather than clarify the profound physical picture. To circumvent all those limitations of the conventional method and further advance this seminal field, new concepts and techniques have to be introduced. 

Here we investigate the problem of electromagnetic scatterings from a different perspective of engineered QNM~\cite{LALANNE__LaserPhotonicsRev._Light,GRAS_2020_J.Opt.Soc.Am.A_Nonuniqueness} excitations and interferences. Scattered waves by the particles can be expanded into  QNM radiations, and thus relative amplitudes and phases among them would decide the scattering patterns. Relying on the principle of reciprocity~\cite{CALOZ_2018_Phys.Rev.Applied_Electromagnetic}, we manage to provide full geometric representations for the excitation coefficients and discover the geometric phase 
(Pancharatnam-Berry phase)~\cite{PANCHARATNAM_1956_ProcIndianAcadSci_Generalized,BERRY_1984_Proc.R.Soc.A_Quantal,berry_adiabatic_1987,BERRY_2024_Opt.PhotonicsNews_GeometricPhase} of QNMs that varies with varying incident polarizations. We further exploit the unveiled geometric phase and the also incident polarization-dependent excitation amplitudes for scattering intensity and polarization manipulations, such as eliminating total and directional scatterings and designing directions of polarization singularities. Continuous geometric phases from $0$-$2\pi$ are directly manifest through scattering variations, even in the rather elementary configuration of an individual particle scattering plane waves of varying polarizations.  Our work has essentially provided an exhaustive framework for the calculation of geometric phase in scattering systems, which can potentially accelerate both fundamental explorations and practical applications relying on scatterings, not only of electromagnetic waves, but also of waves of other forms where the geometric phase would generically emerge.

The eigenmodes supported by an arbitrary open scattering system, also termed as QNMs, can be directly calculated and they generally feature finite Q-factors and complex eigenfrequencies~\cite{LALANNE__LaserPhotonicsRev._Light,GRAS_2020_J.Opt.Soc.Am.A_Nonuniqueness}. When the system is excited by an external source, its scattered (near and far) fields can be expanded as:
\begin{equation}
\label{expansion}
\mathbf{E}_{\rm{sca}}(\mathbf{{r}})=\sum_m \alpha_{m} {\mathbf{E}}_{m}(\mathbf{{r}}),  
\end{equation}
where ${\mathbf{E}}_{m}(\mathbf{{r}})$ denotes the radiation of the $m^{th}$ QNM and $\alpha_{m}$ is the complex expanding (excitation) coefficient. In the far field, both $\mathbf{E}_{\rm{sca}}(\mathbf{{r}})$ and  ${\mathbf{E}}_m(\mathbf{{r}})$ are transverse, and thus Eq.~(\ref{expansion}) can be reformulated as $\rm{E}_{\rm{sca}}(\mathbf{\hat{r}})\mathbf{{J}}_{sca}(\mathbf{\hat{r}})=\sum \alpha_{m}(\omega) {{\rm{E}}}_m(\mathbf{\hat{r}}){\mathbf{{J}}}_{m}(\mathbf{\hat{r}})$. Here $\mathbf{\hat{r}}$ is unit direction vector $\mathbf{\hat{r}}=\mathbf{r}/|\mathbf{r}|$; $\rm{E}_{\rm{sca}}$ and $\rm{E}_m$ are field amplitudes; $\mathbf{{J}}_{\rm{sca}}(\mathbf{\hat{r}})$ and $\mathbf{{J}}_m(\mathbf{\hat{r}})$ are the corresponding unit Jones (row) vectors~\cite{YARIV_2006__Photonics}. Both scattering intensity and polarization distributions are dictated by the relative amplitudes and phases among the excitation coefficients $\alpha_{m}$, which are then decided by the incident wave (initial condition). Conventional calculations of $\alpha_{m}$ rely on near-field integrations that require accurate field distributions at every point inside the scattering bodies~\cite{LALANNE__LaserPhotonicsRev._Light,GRAS_2020_J.Opt.Soc.Am.A_Nonuniqueness,CHEN_Phys.Rev.Lett._Extremize}. 

\begin{figure}[tp]
\centerline{\includegraphics[width=9cm]{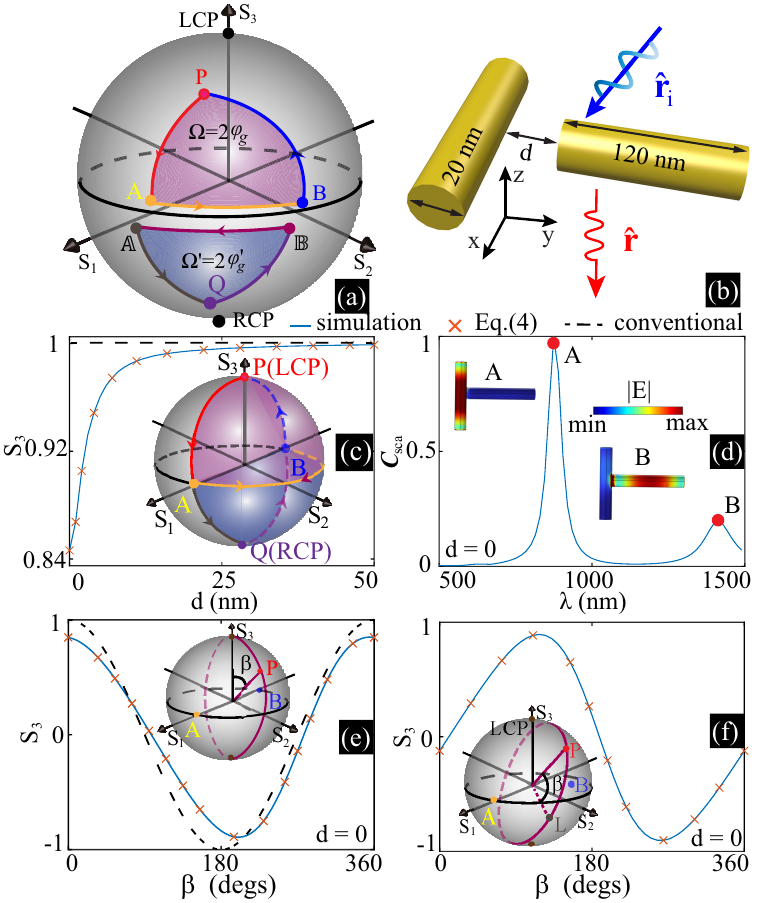}} \caption{\small (a) Poincaré sphere on which P (Q), A ($\mathbb{A}$) and B ($\mathbb{B}$) represent respectively the incident (projected) polarization and polarizations of the mode radiations opposite to the incident direction (along the scattering direction $\mathbf{\hat{r}}$). The geometric phases during incident coupling (scattered field projection) is half solid angle of geodesic ciruit  PABA ($\mathbb{A}$Q$\mathbb{B}$$\mathbb{A}$): $\varphi_g=\frac{1}{2}\Omega$ ($\varphi_g^\prime=\frac{1}{2}\Omega^\prime$). (b) A pair of perpendicular gold cylinders. Dependence of $S_3$ for scatterings along -\textbf{z} on inter-particle distance $d$ (c), and incident polarizations transversing a great circle parameterized by $\beta$ [(e) perpendicular incidence with $\mathbf{\hat{r}}_{\rm{i}}$ along -\textbf{z}; (f) tilted incidence with $\mathbf{\hat{r}}_{\rm{i}}\perp\mathbf{x}$ and $\angle\mathbf{\hat{r}}_{\rm{i}}\mathbf{y}=\pi/4$]. (c) Scattering spectra for LCP incidence along  -\textbf{z}, where the near fields of the two QNMs excited are shown as insets. In (d)-(f) $d=0$ and in (c), (e)  \&(f) the incident wavelength $\lambda=0.8~\mu$m and A (B) locates at $S_1=1$ ($S_1=-1$) except that in (f) B locates at (-0.92, 0.316, -0.236).}
\label{fig1}
\end{figure}

For an incident (along $\mathbf{\hat{r}}_{\rm{i}}$) plane wave of wavelength $\lambda$ and polarization Jones vector $\mathbf{{J}}_{\rm{i}}$, if the scattering system is reciprocal, it is recently revealed that $\alpha_{m}$ can be alternatively calculated in the far field~\cite{CHEN_Phys.Rev.Lett._Extremize}:
\begin{equation}
\label{expansion-coefficient-jones}
\alpha_m \propto \mathbf{J_{\rm{i}}}\mathbf{J}_m^{\dagger}(-\mathbf{\hat{r}}_{\rm{i}}),
\end{equation}
where $\dagger$  denotes combined operations of complex conjugate and transpose, and thus $\mathbf{J}_m^{\dagger}(-\mathbf{\hat{r}}_{\rm{i}})$ is the corresponding column Jones vector for the QNM radiation along  $-\mathbf{\hat{r}}_{\rm{i}}$ (opposite to the incident direction). To reveal the core principles, we start with the simplest scenario of two-QNM (denoted by modes A and B) excitations and the framework established can be naturally generalized to deal with multi-mode cases. To compact the notations, we simplify $\mathbf{J}_{\rm{A,B}}(-\mathbf{\hat{r}}_{\rm{i}})$ as $\mathbf{J}_{\rm{A,B}}$ and the Jones vector for radiation along an arbitrary scattering direction $\mathbf{\hat{r}}$ as $\mathbb{J}_{\rm{A,B}}$, with the direction vector $\mathbf{\hat{r}}$ suppressed for both Jones vectors and field vectors. Then according to Eq.~(\ref{expansion}), the scattered far field is~\cite{RAMACHANDRAN_1961,PANCHARATNAM_1975__Collecteda,berry_adiabatic_1987}:
\begin{equation}
\label{expansion-simplified-2}
\mathbf{E}_{\rm{sca}} \propto |\mathbf{J_{\rm{i}}}\mathbf{J}^{\dagger}_{\rm{A}}|~{\mathbf{E}}_{\rm{A}}\exp({i\varphi_A})+|\mathbf{J_{\rm{i}}}\mathbf{J}^{\dagger}_{\rm{B}}|~{\mathbf{E}}_{\rm{B}} \exp({i\varphi_B}). 
\end{equation}
To geometrize the relative excitation amplitude and phase,  we map the Jones vectors of $\mathbf{J_{\rm{i}}}$, $\mathbf{J}_{\rm{A}}$ ($\mathbb{J}_{\rm{A}}$) and $\mathbf{J}_{\rm{B}}$ ($\mathbb{J}_{\rm{B}}$) to points P, A($\mathbb{A}$), B($\mathbb{B}$) on the Poincaré sphere~\cite{YARIV_2006__Photonics,RAMACHANDRAN_1961,PANCHARATNAM_1975__Collecteda,berry_adiabatic_1987} of unit radius and parameterized by three Stokes parameters $S_{1,2,3}$ [shown in Fig.~\ref{fig1}(a)]. Then Eq.~(\ref{expansion-simplified-2}) has a pure geometric form~\cite{Supplemental_Material}:
\begin{equation}
\label{expansion-simplified}
\mathbf{E}_{\rm{sca}} \propto \cos\left(\frac{1}{2}\arc{PA}\right)~{\mathbf{E}}_{\rm{A}}+\cos\left(\frac{1}{2}\arc{PB}\right)~{\mathbf{E}}_{\rm{B}} \exp({i\varphi_g}). 
\end{equation}
Here $\arc{PA}$ ($\arc{PB}$) denotes the length of the great arc (shorter segment $\in[0, \pi]$) connecting PA (PB);  for example, when the incident polarization is orthogonal to that of mode A  along $-\mathbf{\hat{r}}_{\rm{i}}$ ($\mathbf{J_{\rm{i}}}\mathbf{J}^{\dagger}_{\rm{A}}=0$), P and A are antipodal points ($\arc{PA}=\pi$) and thus mode A would not be excited [$\cos(\frac{1}{2}\arc{PA})=0$], being consistent with the special scenario of single-mode excitations~\cite{CHEN_Phys.Rev.Lett._Extremize,WEN_2024_Laser&PhotonicsReviews_Momentum}. The phase contrast $\varphi_g=\varphi_B-\varphi_A$ is a geometric phase~\cite{Supplemental_Material}: $\varphi_g=\frac{1}{2}\Omega$, where $\Omega$ denotes the solid angle of the great-arc (geodesic) circuit PABP [see Fig.~\ref{fig1}(a); $\Omega$ is positive (negative) for counter-clockwise (clockwise) circuit viewed above~\cite{PANCHARATNAM_1956_ProcIndianAcadSci_Generalized,BERRY_1984_Proc.R.Soc.A_Quantal,berry_adiabatic_1987,BERRY_2024_Opt.PhotonicsNews_GeometricPhase}.

Relying on Eq.~(\ref{expansion-simplified}), the scattering intensity $I_{\rm{sca}}=|\mathbf{E}_{\rm{sca}}|^2$ and polarization along arbitrary directions can be calculated. When the scattered field $\mathbf{E}_{\rm{sca}}$ is projected  to a specific polarization [point Q on the Poincaré sphere with Jones vector  $\mathbf{J}_{\rm{Q}}$; see Fig.~\ref{fig1}(a)], such as being analyzed by a polarizer, we obtain:
\begin{equation}
\label{expansion-simplified3}
\begin{split}
\mathbf{E}_{\rm{sca}}^Q\propto \mathbf{J}_{\rm{Q}} \bigg[\cos\left(\frac{1}{2}\arc{PA}\right)\cos\left(\frac{1}{2}\mathbf{\overset{\large\frown}{\mathbb{A}\mathrm{Q}}}\right){{\mathrm{E}}}_{\rm{A}}+\\
\cos\left(\frac{1}{2}\arc{PB}\right)\cos\left(\frac{1}{2}\mathbf{\overset{\large\frown}{\mathbb{B}\mathrm{Q}}}\right){{\mathrm{E}}}_{\rm{B}} \exp[{i(\varphi_g+{\varphi}_g^\prime)}\bigg]. 
\end{split}
\end{equation}
Here the extra geometric phase term comes from the projection $\varphi_g^\prime=\frac{1}{2}\Omega^\prime$, where $\Omega^\prime$ denotes the solid angle of the geodesic circuit $\mathbb{A}$Q$\mathbb{B}$$\mathbb{A}$ [see Fig.~\ref{fig1}(a)]~\cite{RAMACHANDRAN_1961,PANCHARATNAM_1975__Collecteda,berry_adiabatic_1987}. The total geometric phase $\varphi_G=\varphi_g+\varphi_g^\prime$ has to be calculatedly separately, unless A (B) and $\mathbb{A}$ ($\mathbb{B}$) overlap (mode radiation polarizations along $-{\mathbf{\hat{r}}}_{\rm{i}}$ and $\mathbf{\hat{r}}$ are identical), when it is half the solid angle of the whole geodesic circuit PAQBP.

We emphasize that the theoretical framework that has been so far elaborated on is fully based on the QNMs of open photonic systems, and thus is generically applicable to scattering bodies (individual or clustered) of arbitrary geometric shapes and optical parameters, requiring only the reciprocity that guarantees the validity of Eq.~(\ref{expansion-coefficient-jones}). Furthermore, the geometric phase involved consist of two parts that have different origins: $\varphi_g$ originates from  the coupling of incident polarization (P) onto the QNM radiation polarizations (A and B) along $-\mathbf{\hat{r}}_{\rm{i}}$ [Eq.~(\ref{expansion-coefficient-jones})]; while $\varphi_g^\prime$ originates from  the projections of QNM radiation polarizations along an arbitrary scattering direction $\mathbf{\hat{r}}$ ($\mathbb{A}$ and $\mathbb{B}$) to any desired polarization of interest (Q). Since the formalisms have put no constraints on $\mathbf{\hat{r}}_{\rm{i}}$ or $\mathbf{\hat{r}}$, our model is applicable to arbitrary incident and scattering directions. For applications that do not involve the scattering polarization projection,  Eq.~(\ref{expansion-simplified}) contains all information and  $\varphi_g^\prime$ will become superfluous.

We now turn to a specific bi-cylinder scattering configuration shown in Fig.~\ref{fig1}(b) to verify our theoretical framework (numerical calculations are performed using COMSOL Multiphysics throughout this work). The cylinders are identical, perpendicular to each other and  consists of gold (effective permittivity fitted from data in Ref.~\cite{Johnson1972_PRB}). The individual cylinder supports an electric dipole (ED) at the spectral regime of interest~\cite{Supplemental_Material}.  Similar structures consisting of optical items with spatially varying orientations are widely employed for various photonic functionalities based on geometric phase~\cite{LI_Nat.Rev.Mater._nonlinear_2017,COHEN_2019_NatRevPhys_Geometric,JISHA_2021_LaserPhotonicsRev._Geometric,GUO_2022_PI_Classical}. The mostly widely discussed scattering configuration is shining, for example, left-handed circularly-polarized (LCP and $S_3=1$) waves along -\textbf{z} and then project the forward scattered waves onto right-handed circularly-polarized (RCP;  $S_3=-1$) states. The conventional pictorial representation of the geometric phase ($\varphi_G=\pi$) is shown in the inset of Fig.~\ref{fig1}(c). From the perspective of our theoretical framework, such a representation is a reduced approximation of our model with the following two requirements: (i) each consisting cylinder supports an ED, with radiation polarizations along $\pm$\textbf{z} being both linear and parallel to the cylinder orientations; (ii) the couplings between the cylinders are negligible, ensuring that the EDs supported by both cylinders are also the QNMs of the bi-cylinder system. (i) \& (ii) leads to overlapped A (B) with $\mathbb{A}$ ($\mathbb{B}$) on the equator of the Poincaré sphere, and thus the conventional representation [inset of Fig.~\ref{fig1}(c)]  is merely a special scenario of the our general representation in Fig.~\ref{fig1}(a). However, when the modes supported are not EDs and/or the coupling is not negligible, the conventional approach to calculate geometric phase relying on structural orientations would break down. 

To showcase the superiority of our model, we show in Fig.~\ref{fig1}(c) the evolution of the polarization (in terms of $S_3$) for the forward scatterings along -\textbf{z} with fixed LCP incidence while varying inter-cylinder distance $d$. For the conventional approach, the RCP components cancel each other due to destructive interference ($\varphi_G=\pi$) and thus the forward scattered waves maintain to be LCP with fixed $S_3=1$. Nevertheless, as is manifest in Fig.~\ref{fig1}(c), with decreasing $d$ with stronger inter-cylinder couplings, the forward scattered waves will contain both RCP and LCP components, where our model still works. For the extreme case of $d=0$, we further show in  Fig.~\ref{fig1}(d) the scattering cross section ($C_\mathrm{sca}$) spectra to confirm the excitations of two QNMs (LCP incidence), near fields of which are shown as insets. The QNM radiations along the opposite incident direction (+\textbf{z}) are linear~\cite{Supplemental_Material} and their positions are indicated in the inset of Fig.~\ref{fig1}(e), where we show the forward scattering polarization evolutions with varying incident polarizations. Here P locates on a great circle parameterized by $\beta$ ($\beta=0$ for LCP incidence) and  equally bisects \arc{AB} (\arc{PA}=\arc{PB}). 
We then tilt the incident direction ($\mathbf{\hat{r}}_{\rm{i}}\perp\mathbf{x}$ and $\angle\mathbf{\hat{r}}_{\rm{i}}\mathbf{y}=\pi/4$) and show the scattering polarization along +\textbf{z} in Fig.~\ref{fig1}(f) for P transversing another great circle equally bisecting \arc{AB} (see the inset; $\beta=0$ corresponds to state L that locates on \arc{AB}).  Since radiation of mode B opposite to the incident direction are not linearly polarized anymore, B does not locate on the equator anymore. For both scenarios of perpendicular and tilt incidences [Figs.~\ref{fig1}(e) and \ref{fig1}(f)], results from our model agree perfectly well with the simulation results. The results for the non-perpendicular bi-cylinder configuration and for another classical structure consisting of a pair of twisted split-ring resonators~\cite{Supplemental_Material} further confirm the validity of our theoretical model.

\begin{figure}[tp]
\centerline{\includegraphics[width=9cm]{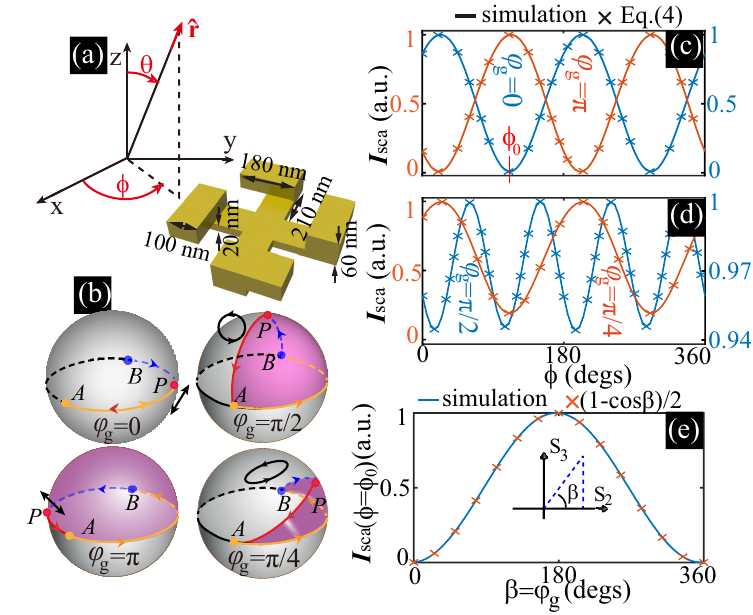}} \caption{\small (a) A gold particle with $4$-fold rotation symmetry and a spherical coordinate system parameterized by $\mathbf{r}=(r,~\theta,~\phi)$. Scattering intensity distributions on the $\mathbf{x}$-$\mathbf{y}$ plane with P locating at ($0$, $1$, $0$) and ($0$, $-1$, $0$) for (c), which correspond to two orthogonal incident linear polarizations [polarized along $\phi=\pi/4$ ($S_2=1$) and polarized along $\phi=3\pi/4$ ($S_2=-1$)];  locating at ($0$, $0$, $1$) and ($0$, $\sqrt{2}/2$, $\sqrt{2}/2$) for (d), which correspond to  LCP incident and elliptic incident polarization. The geometric phases are respectively $\varphi_g=0,\pi,\pi/2$ and $\pi/4$, as shown in (b).  In (c) one direction of  zero scattering $\phi=\phi_0\approx113^\circ$ is specified. (e) The dependence of $I_{\rm{sca}}(\phi=\phi_0)$ on $\beta$ (see the inset). In (c)-(e) the incident wavelength $\lambda=2.34~\mu$m.}
\label{fig2}
\end{figure}

In our theoretical framework, the calculation of the geometric phase has nothing to do with orientations of the structures, and thus our model is applicable to individual particles without any preferred orientations. The gold particle shown in Fig.~\ref{fig2}(a) exhibits four-fold rotation symmetry that secures a pair of degenerate QNMs~\cite{Supplemental_Material}. We shine plane waves along +$\mathbf{z}$  and track scattering intensity distributions on the $\mathbf{x}$-$\mathbf{y}$ plane. The radiations of the two QNMs along -$\mathbf{z}$ (opposite to the incident direction) are almost linearly polarized parallel to $\mathbf{x}$ and $\mathbf{y}$ [see points A and B in Fig.~\ref{fig2}(b)]~\cite{Supplemental_Material}. The normalized scattering intensity distributions [parameterized by the azimuthal angle $\phi$ as shown in Fig.~\ref{fig2}(a)] are shown in Figs.~\ref{fig2}(c) and \ref{fig1}(d), for four incident polarizations with the corresponding geometric phase [Fig.~\ref{fig2}(b)]  $\varphi_g\approx0,~\pi$ (two perpendicular linear polarizations) and $\varphi_g\approx\pi/2,~\pi/4$ (circular and elliptical polarizations), respectively. In Fig.~\ref{fig2}(c), the geometric phase $\pi$ is directly manifest through the two distinct curves and we note that our demonstration (with an individual scattering particle) of such classical geometric phase is even simpler and more direct than the earliest ones by Fresnel-Arago and Hamilton-Lloyd~\cite{BERRY_2024_Opt.PhotonicsNews_GeometricPhase}.  As is shown in Fig.~\ref{fig2}(c), for $\varphi_g\approx0$ the scattering along the direction $\phi=\phi_0\approx113^\circ$ is zero. Along this direction, according to Eq.~(\ref{expansion-simplified}) we have $\mathbf{E}_{\rm{A}}\approx-\mathbf{E}_{\rm{B}}$ since $\arc{PA}\approx\arc{PB}\approx\pi/2$ and $\varphi_g\approx0$ [Fig.~\ref{fig2}(b)].  With fixed $\arc{PA}\approx\arc{PB}$ while changing $\varphi_g$, we then have $\mathbf{E}_{\rm{sca}} \propto 1-\exp({i\varphi_g})$ and thus scattering intensity along this direction: $I_{\rm{sca}}  \propto  |\mathbf{E}_{\rm{sca}}|^2  \propto 1-\cos(\varphi_g)$. We then show the evolutions of $I_{\rm{sca}}(\phi=\phi_0)$ on such a polarization great circle ($S_1=0$) parameterized by $\beta\in [0,2\pi]$, as shown in the inset in Fig.~\ref{fig2}(e). Obviously the solid angle $\Omega=2\beta$ and $\varphi_g=\beta$ [Fig.~\ref{fig2}(b)], and thus the evolution observe the relation $I_{\rm{sca}}(\phi=\phi_0)\propto 1-\cos(\beta)$, which agrees perfectly with the numerical results included in Fig.~\ref{fig2}(e).

\begin{figure}[tp]
\centerline{\includegraphics[width=9cm]{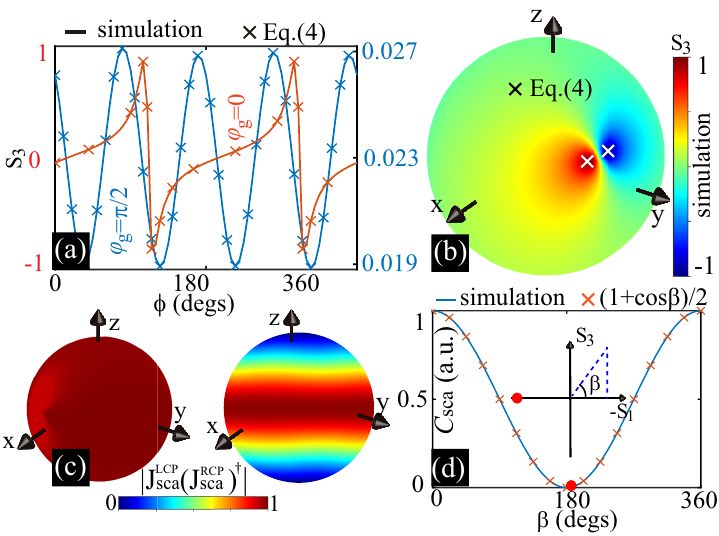}} \caption{\small (a) $S_3$ distributions on the $\mathbf{x}$-$\mathbf{y}$ plane with P locating at ($0$, $0$, $1$) and ($0$, $1$, $0$), with $\varphi_g=\pi/2$ and $0$, receptively. (b) $S_3$ distributions on the whole scattering momentum sphere for P locating at ($0$, $1$, $0$), on which the predicted directions of circular polarization singularities are marked by crosses. (c) Distributions of $|\mathbf{J}_{\rm{sca}}^{\rm{LCP}}(\mathbf{J}_{\rm{sca}}^{\rm{RCP}})^\dagger|$  for waves incident along $\mathbf{\hat{r}}_{\rm{i}}=-\mathbf{\hat{r}}_{\rm{A=B}}$ (left) and $\mathbf{\hat{r}}_{\rm{i}} =+\mathbf{z}$ (right). (d) Variations of $C_{\rm{sca}}$ for P transversing a great circle ($S_2=0$), covering matched polarization ($\beta=0$),  orthogonal polarization (marked; $\beta=\pi$), LCP and RCP ($\beta=\pi/2$ and $3\pi/2$). The incident wavelength $\lambda=2.34~\mu$m.}
\label{fig3}
\end{figure}

The polarization distributions on the $\mathbf{x}$-$\mathbf{y}$ plane are further shown in Fig.~\ref{fig3}(a), for linear polarization ($S_2=1$ and $\varphi_g\approx0$) and LCP ($S_3=1$ and $\varphi_g\approx\pi/2$) incidences. According to Eq.~(\ref{expansion-simplified}), the locations of circularly-polarized scatterings ($S_3=\pm1$; circular polarization singularities~\cite{GBUR_2016__Singular}) can be directly predicted and even designed by selecting proper incident polarization and directions. We show the scattering polarization distributions (simulated results) on the whole scattering momentum sphere in Fig.~\ref{fig3}(b) for the linear polarization incidence ($S_2=1$) and the predicted locations of polarization singularities are also marked by crosses, agreeing well with the numerical calculations. 

Also according to Eq.~(\ref{expansion-simplified}), there is a rather interesting scenario of overlapped A and B: directions along which the radiation polarizations for both QNMs are the same. For waves incident opposite to those directions, we have  $\cos(\frac{1}{2}\arc{PA})=\cos(\frac{1}{2}\arc{PB})$ and $\varphi_g=0$ [see Fig.~\ref{fig1}(a)]. This results in fixed scattering polarization along any direction, irrespective of varying incident polarizations.  For the marked direction ($\phi=\phi_0$) in Fig.~\ref{fig2}(a) where the scattering is zero, since both modes are excited, along this direction QNMs radiation polarizations must be the same or they would not cancel each other. We denote this direction as $\mathbf{\hat{r}}_{\rm{A=B}}$ and shine opposite to it ($\mathbf{\hat{r}}_{\rm{i}}=-\mathbf{\hat{r}}_{\rm{A=B}}$) LCP and RCP waves. We then track the scattering polarization variation on the whole momentum sphere through the parameter $|\mathbf{J}_{\rm{sca}}^{\rm{LCP}}(\mathbf{J}_{\rm{sca}}^{\rm{RCP}})^\dagger|$, where $|\mathbf{J}_{\rm{sca}}^{\rm{LCP}}(\mathbf{J}_{\rm{sca}}^{\rm{RCP}})^\dagger|=1$ means the scattering polarization does not change for RCP and LCP incidences [see Fig.~\ref{fig3}(c)]. For comparison, we also shine circularly-polarized waves along +$\mathbf{z}$ (opposite to which the QNM radiations are distinct and thus A and B do not overlap) and show its $|\mathbf{J}_{\rm{sca}}^{\rm{LCP}}(\mathbf{J}_{\rm{sca}}^{\rm{RCP}})^\dagger|$ distribution in Fig.~\ref{fig3}(c), confirming that for general incident directions the scattering polarizations are dependent on incident polarizations. 

Another interesting property for overlapped A and B is that the incident polarization can be tuned to be orthogonal to that of both QNMs  [$\cos(\frac{1}{2}\arc{PA})=\cos(\frac{1}{2}\arc{PB})=0$; $\mathbf{E}_{\rm{sca}}=0$ in Eq.~(\ref{expansion-simplified})]. For such an incident polarization, neither QNM would be excited and thus the particle would become invisible. For the same incident direction ($\mathbf{\hat{r}}_{\rm{i}}=-\mathbf{\hat{r}}_{\rm{A=B}}$), we show the evolutions of scattering cross sections with varying incident polarizations in Fig.~\ref{fig3}(d). The incident polarization covers a great circle, covering matched polarization ($\beta=\arc{PA}=\arc{PB}=0$), LCP ($\beta=\pi/2$), orthogonal polarization ($\beta=\arc{PA}=\arc{PB}=\pi$). According to Eq.~(\ref{expansion-simplified}) with $\varphi_g=0$ and $\arc{PA}=\arc{PB}=\beta$: $\mathbf{E}_{\rm{sca}} \propto ({\mathbf{E}}_{\rm{A}}+{\mathbf{E}}_{\rm{B}})\cos(\beta/2)$; that is, scattering along any direction and thus also the scattering cross section would satisfy $I_{\rm{sca}}, C_{\rm{sca}}~\propto \cos^2(\beta/2) =[1+\cos(\beta)]/2$, which is verified by Fig.~\ref{fig3}(d).  We note that the scattering evolutions in Fig.~\ref{fig2}(e) are driven by changing $\varphi_g$ with fixed $\arc{PA}=\arc{PB}=\pi/2$ [as is also the case for those in Figs.~\ref{fig1}(e) and \ref{fig1}(f)], while those in Fig.~\ref{fig3}(d) driven by changing $\arc{PA}=\arc{PB}=\beta$, with fixed $\varphi_g=0$.

In conclusion, we have unveiled the hidden excitation geometric phases among QNMs of scatterers and reveal how they drive scattering evolutions on the incident Poincaré sphere. The geometric phase can be exploited to efficiently manipulate the scatterings, such as scattering eliminations and  polarization singularity (zero and circularly-polarized scatterings) generations. For the general scenario of more than two QNMs being simultaneously excited, the relative amplitude and phase among any two QNMs can be calculated using our model and then calculations for interferences among all QNMs become routine. This means that the theoretical framework we have constructed is generic and widely applicable. We have essentially established a profoundly comprehensive framework to calculate geometric phases in scattering systems, and unlocked an extra hidden dimension for electromagnetic scattering manipulations. Similar dimensions might be uncovered for waves of other forms for which geometric phase is generic and ubiquitous, providing new flexibilities for many physics and interdisciplinary branches that are related to wave scatterings.

\emph{Acknowledgment}: This work is supported by National Natural Science Foundation of China (Grant No. 11874426 and 61405067), Outstanding
Young Researcher Scheme of Hunan Province (2024JJ2056), and several Researcher Schemes of National University of Defense Technology. W. L. acknowledges many illuminating correspondences with Sir Michael Berry and this paper was deliberately submitted on the 40$^{\mathrm{th}}$ birthday of his monumental paper on geometric phase. 



\begin{thebibliography}{39}%
\makeatletter
\providecommand \@ifxundefined [1]{%
 \@ifx{#1\undefined}
}%
\providecommand \@ifnum [1]{%
 \ifnum #1\expandafter \@firstoftwo
 \else \expandafter \@secondoftwo
 \fi
}%
\providecommand \@ifx [1]{%
 \ifx #1\expandafter \@firstoftwo
 \else \expandafter \@secondoftwo
 \fi
}%
\providecommand \natexlab [1]{#1}%
\providecommand \enquote  [1]{``#1''}%
\providecommand \bibnamefont  [1]{#1}%
\providecommand \bibfnamefont [1]{#1}%
\providecommand \citenamefont [1]{#1}%
\providecommand \href@noop [0]{\@secondoftwo}%
\providecommand \href [0]{\begingroup \@sanitize@url \@href}%
\providecommand \@href[1]{\@@startlink{#1}\@@href}%
\providecommand \@@href[1]{\endgroup#1\@@endlink}%
\providecommand \@sanitize@url [0]{\catcode `\\12\catcode `\$12\catcode
  `\&12\catcode `\#12\catcode `\^12\catcode `\_12\catcode `\%12\relax}%
\providecommand \@@startlink[1]{}%
\providecommand \@@endlink[0]{}%
\providecommand \url  [0]{\begingroup\@sanitize@url \@url }%
\providecommand \@url [1]{\endgroup\@href {#1}{\urlprefix }}%
\providecommand \urlprefix  [0]{URL }%
\providecommand \Eprint [0]{\href }%
\providecommand \doibase [0]{https://doi.org/}%
\providecommand \selectlanguage [0]{\@gobble}%
\providecommand \bibinfo  [0]{\@secondoftwo}%
\providecommand \bibfield  [0]{\@secondoftwo}%
\providecommand \translation [1]{[#1]}%
\providecommand \BibitemOpen [0]{}%
\providecommand \bibitemStop [0]{}%
\providecommand \bibitemNoStop [0]{.\EOS\space}%
\providecommand \EOS [0]{\spacefactor3000\relax}%
\providecommand \BibitemShut  [1]{\csname bibitem#1\endcsname}%
\let\auto@bib@innerbib\@empty
\bibitem [{\citenamefont {Bohren}\ and\ \citenamefont
  {Huffman}(1983)}]{Bohren1983_book}%
  \BibitemOpen
  \bibfield  {author} {\bibinfo {author} {\bibfnamefont {C.~F.}\ \bibnamefont
  {Bohren}}\ and\ \bibinfo {author} {\bibfnamefont {D.~R.}\ \bibnamefont
  {Huffman}},\ }\href@noop {} {\emph {\bibinfo {title} {Absorption and
  Scattering of Light by Small Particles}}}\ (\bibinfo  {publisher} {Wiley},\
  \bibinfo {year} {1983})\BibitemShut {NoStop}%
\bibitem [{\citenamefont {Kuznetsov}\ \emph {et~al.}(2016)\citenamefont
  {Kuznetsov}, \citenamefont {Miroshnichenko}, \citenamefont {Brongersma},
  \citenamefont {Kivshar},\ and\ \citenamefont
  {Luk'yanchuk}}]{KUZNETSOV_Science_optically_2016}%
  \BibitemOpen
  \bibfield  {author} {\bibinfo {author} {\bibfnamefont {A.~I.}\ \bibnamefont
  {Kuznetsov}}, \bibinfo {author} {\bibfnamefont {A.~E.}\ \bibnamefont
  {Miroshnichenko}}, \bibinfo {author} {\bibfnamefont {M.~L.}\ \bibnamefont
  {Brongersma}}, \bibinfo {author} {\bibfnamefont {Y.~S.}\ \bibnamefont
  {Kivshar}},\ and\ \bibinfo {author} {\bibfnamefont {B.}~\bibnamefont
  {Luk'yanchuk}},\ }\bibfield  {title} {\bibinfo {title} {Optically resonant
  dielectric nanostructures},\ }\href@noop {} {\bibfield  {journal} {\bibinfo
  {journal} {Science}\ }\textbf {\bibinfo {volume} {354}},\ \bibinfo {pages}
  {aag2472} (\bibinfo {year} {2016})}\BibitemShut {NoStop}%
\bibitem [{\citenamefont {Liu}\ and\ \citenamefont
  {Kivshar}(2018)}]{LIU_2018_Opt.Express_Generalized}%
  \BibitemOpen
  \bibfield  {author} {\bibinfo {author} {\bibfnamefont {W.}~\bibnamefont
  {Liu}}\ and\ \bibinfo {author} {\bibfnamefont {Y.~S.}\ \bibnamefont
  {Kivshar}},\ }\bibfield  {title} {\bibinfo {title} {Generalized {{Kerker}}
  effects in nanophotonics and meta-optics {{[Invited]}}},\ }\href@noop {}
  {\bibfield  {journal} {\bibinfo  {journal} {Opt. Express}\ }\textbf {\bibinfo
  {volume} {26}},\ \bibinfo {pages} {13085} (\bibinfo {year}
  {2018})}\BibitemShut {NoStop}%
\bibitem [{\citenamefont {Krasnok}\ \emph {et~al.}(2019)\citenamefont
  {Krasnok}, \citenamefont {Baranov}, \citenamefont {Li}, \citenamefont {Miri},
  \citenamefont {Monticone},\ and\ \citenamefont
  {Al{\'u}}}]{KRASNOK_2019_Adv.Opt.Photon.AOP_Anomalies}%
  \BibitemOpen
  \bibfield  {author} {\bibinfo {author} {\bibfnamefont {A.}~\bibnamefont
  {Krasnok}}, \bibinfo {author} {\bibfnamefont {D.}~\bibnamefont {Baranov}},
  \bibinfo {author} {\bibfnamefont {H.}~\bibnamefont {Li}}, \bibinfo {author}
  {\bibfnamefont {M.-A.}\ \bibnamefont {Miri}}, \bibinfo {author}
  {\bibfnamefont {F.}~\bibnamefont {Monticone}},\ and\ \bibinfo {author}
  {\bibfnamefont {A.}~\bibnamefont {Al{\'u}}},\ }\bibfield  {title} {\bibinfo
  {title} {Anomalies in light scattering},\ }\href
  {https://doi.org/10.1364/AOP.11.000892} {\bibfield  {journal} {\bibinfo
  {journal} {Adv. Opt. Photon.}\ }\textbf {\bibinfo {volume} {11}},\ \bibinfo
  {pages} {892} (\bibinfo {year} {2019})}\BibitemShut {NoStop}%
\bibitem [{\citenamefont {Kivshar}(2022)}]{KIVSHAR_2022_NanoLett._Rise}%
  \BibitemOpen
  \bibfield  {author} {\bibinfo {author} {\bibfnamefont {Y.}~\bibnamefont
  {Kivshar}},\ }\bibfield  {title} {\bibinfo {title} {The {{Rise}} of
  {{Mie-tronics}}},\ }\href {https://doi.org/10.1021/acs.nanolett.2c00548}
  {\bibfield  {journal} {\bibinfo  {journal} {Nano Lett.}\ }\textbf {\bibinfo
  {volume} {22}},\ \bibinfo {pages} {3513} (\bibinfo {year}
  {2022})}\BibitemShut {NoStop}%
\bibitem [{\citenamefont {Altug}\ \emph {et~al.}(2022)\citenamefont {Altug},
  \citenamefont {Oh}, \citenamefont {Maier},\ and\ \citenamefont
  {Homola}}]{ALTUG_2022_Nat.Nanotechnol._Advances}%
  \BibitemOpen
  \bibfield  {author} {\bibinfo {author} {\bibfnamefont {H.}~\bibnamefont
  {Altug}}, \bibinfo {author} {\bibfnamefont {S.-H.}\ \bibnamefont {Oh}},
  \bibinfo {author} {\bibfnamefont {S.~A.}\ \bibnamefont {Maier}},\ and\
  \bibinfo {author} {\bibfnamefont {J.}~\bibnamefont {Homola}},\ }\bibfield
  {title} {\bibinfo {title} {Advances and applications of nanophotonic
  biosensors},\ }\href {https://doi.org/10.1038/s41565-021-01045-5} {\bibfield
  {journal} {\bibinfo  {journal} {Nat. Nanotechnol.}\ }\textbf {\bibinfo
  {volume} {17}},\ \bibinfo {pages} {5} (\bibinfo {year} {2022})}\BibitemShut
  {NoStop}%
\bibitem [{\citenamefont {Fan}\ and\ \citenamefont
  {Li}(2022)}]{FAN_2022_Nat.Photon._Photonics}%
  \BibitemOpen
  \bibfield  {author} {\bibinfo {author} {\bibfnamefont {S.}~\bibnamefont
  {Fan}}\ and\ \bibinfo {author} {\bibfnamefont {W.}~\bibnamefont {Li}},\
  }\bibfield  {title} {\bibinfo {title} {Photonics and thermodynamics concepts
  in radiative cooling},\ }\href {https://doi.org/10.1038/s41566-021-00921-9}
  {\bibfield  {journal} {\bibinfo  {journal} {Nat. Photon.}\ }\textbf {\bibinfo
  {volume} {16}},\ \bibinfo {pages} {182} (\bibinfo {year} {2022})}\BibitemShut
  {NoStop}%
\bibitem [{\citenamefont {Liu}\ \emph {et~al.}(2021)\citenamefont {Liu},
  \citenamefont {Liu}, \citenamefont {Shi},\ and\ \citenamefont
  {Kivshar}}]{LIU_ArXiv201204919Phys._Topological}%
  \BibitemOpen
  \bibfield  {author} {\bibinfo {author} {\bibfnamefont {W.}~\bibnamefont
  {Liu}}, \bibinfo {author} {\bibfnamefont {W.}~\bibnamefont {Liu}}, \bibinfo
  {author} {\bibfnamefont {L.}~\bibnamefont {Shi}},\ and\ \bibinfo {author}
  {\bibfnamefont {Y.}~\bibnamefont {Kivshar}},\ }\bibfield  {title} {\bibinfo
  {title} {Topological polarization singularities in metaphotonics},\
  }\href@noop {} {\bibfield  {journal} {\bibinfo  {journal} {Nanophotonics}\
  }\textbf {\bibinfo {volume} {10}},\ \bibinfo {pages} {1469} (\bibinfo {year}
  {2021})}\BibitemShut {NoStop}%
\bibitem [{\citenamefont {Miri}\ and\ \citenamefont
  {Al{\`u}}(2019)}]{MIRI_2019_Science_Exceptionala}%
  \BibitemOpen
  \bibfield  {author} {\bibinfo {author} {\bibfnamefont {M.-A.}\ \bibnamefont
  {Miri}}\ and\ \bibinfo {author} {\bibfnamefont {A.}~\bibnamefont {Al{\`u}}},\
  }\bibfield  {title} {\bibinfo {title} {Exceptional points in optics and
  photonics},\ }\href@noop {} {\bibfield  {journal} {\bibinfo  {journal}
  {Science}\ }\textbf {\bibinfo {volume} {363}},\ \bibinfo {pages} {eaar7709}
  (\bibinfo {year} {2019})}\BibitemShut {NoStop}%
\bibitem [{\citenamefont {Koshelev}\ \emph {et~al.}(2019)\citenamefont
  {Koshelev}, \citenamefont {Bogdanov},\ and\ \citenamefont
  {Kivshar}}]{KOSHELEV_2019_ScienceBulletin_Metaoptics}%
  \BibitemOpen
  \bibfield  {author} {\bibinfo {author} {\bibfnamefont {K.}~\bibnamefont
  {Koshelev}}, \bibinfo {author} {\bibfnamefont {A.}~\bibnamefont {Bogdanov}},\
  and\ \bibinfo {author} {\bibfnamefont {Y.}~\bibnamefont {Kivshar}},\
  }\bibfield  {title} {\bibinfo {title} {Meta-optics and bound states in the
  continuum},\ }\href {https://doi.org/10.1016/j.scib.2018.12.003} {\bibfield
  {journal} {\bibinfo  {journal} {Science Bulletin}\ }\bibinfo {series}
  {{{SPECIAL TOPIC}}: {{Electromagnetic Metasurfaces}}: From {{Concept}} to
  {{Applications}}},\ \textbf {\bibinfo {volume} {64}},\ \bibinfo {pages} {836}
  (\bibinfo {year} {2019})}\BibitemShut {NoStop}%
\bibitem [{\citenamefont {Kang}\ \emph {et~al.}(2023)\citenamefont {Kang},
  \citenamefont {Liu}, \citenamefont {Chan},\ and\ \citenamefont
  {Xiao}}]{KANG_2023_NatRevPhys_Applications}%
  \BibitemOpen
  \bibfield  {author} {\bibinfo {author} {\bibfnamefont {M.}~\bibnamefont
  {Kang}}, \bibinfo {author} {\bibfnamefont {T.}~\bibnamefont {Liu}}, \bibinfo
  {author} {\bibfnamefont {C.~T.}\ \bibnamefont {Chan}},\ and\ \bibinfo
  {author} {\bibfnamefont {M.}~\bibnamefont {Xiao}},\ }\bibfield  {title}
  {\bibinfo {title} {Applications of bound states in the continuum in
  photonics},\ }\href {https://doi.org/10.1038/s42254-023-00642-8} {\bibfield
  {journal} {\bibinfo  {journal} {Nat Rev Phys}\ }\textbf {\bibinfo {volume}
  {5}},\ \bibinfo {pages} {659} (\bibinfo {year} {2023})}\BibitemShut {NoStop}%
\bibitem [{\citenamefont {Doicu}\ \emph {et~al.}(2006)\citenamefont {Doicu},
  \citenamefont {Wriedt},\ and\ \citenamefont {Eremin}}]{DOICU_light_2006}%
  \BibitemOpen
  \bibfield  {author} {\bibinfo {author} {\bibfnamefont {A.}~\bibnamefont
  {Doicu}}, \bibinfo {author} {\bibfnamefont {T.}~\bibnamefont {Wriedt}},\ and\
  \bibinfo {author} {\bibfnamefont {Y.~A.}\ \bibnamefont {Eremin}},\
  }\href@noop {} {\emph {\bibinfo {title} {Light Scattering by Systems of
  Particles: Null-Field Method with Discrete Sources: Theory and Programs}}},\
  Vol.\ \bibinfo {volume} {124}\ (\bibinfo  {publisher} {{Springer}},\ \bibinfo
  {year} {2006})\BibitemShut {NoStop}%
\bibitem [{\citenamefont {Needham}(2021)}]{NEEDHAM__Visuala}%
  \BibitemOpen
  \bibfield  {author} {\bibinfo {author} {\bibfnamefont {T.}~\bibnamefont
  {Needham}},\ }\href@noop {} {\emph {\bibinfo {title} {Visual {{Differential
  Geometry}} and {{Forms}}: {{A Mathematical Drama}} in {{Five Acts}}}}}\
  (\bibinfo  {publisher} {{Princeton University Press}},\ \bibinfo {address}
  {{Princeton}},\ \bibinfo {year} {2021})\BibitemShut {NoStop}%
\bibitem [{\citenamefont {Liu}\ \emph {et~al.}(2014)\citenamefont {Liu},
  \citenamefont {Zhang}, \citenamefont {Lei}, \citenamefont {Ma}, \citenamefont
  {Xie},\ and\ \citenamefont {Hu}}]{Liu2014_ultradirectional}%
  \BibitemOpen
  \bibfield  {author} {\bibinfo {author} {\bibfnamefont {W.}~\bibnamefont
  {Liu}}, \bibinfo {author} {\bibfnamefont {J.}~\bibnamefont {Zhang}}, \bibinfo
  {author} {\bibfnamefont {B.}~\bibnamefont {Lei}}, \bibinfo {author}
  {\bibfnamefont {H.}~\bibnamefont {Ma}}, \bibinfo {author} {\bibfnamefont
  {W.}~\bibnamefont {Xie}},\ and\ \bibinfo {author} {\bibfnamefont
  {H.}~\bibnamefont {Hu}},\ }\bibfield  {title} {\bibinfo {title}
  {Ultra-directional forward scattering by individual core-shell
  nanoparticles},\ }\href@noop {} {\bibfield  {journal} {\bibinfo  {journal}
  {Opt. Express}\ }\textbf {\bibinfo {volume} {22}},\ \bibinfo {pages} {16178}
  (\bibinfo {year} {2014})}\BibitemShut {NoStop}%
\bibitem [{\citenamefont {Jackson}(1998)}]{JACKSON_1998__Classical}%
  \BibitemOpen
  \bibfield  {author} {\bibinfo {author} {\bibfnamefont {J.~D.}\ \bibnamefont
  {Jackson}},\ }\href@noop {} {\emph {\bibinfo {title} {Classical
  {{Electrodynamics Third Edition}}}}},\ \bibinfo {edition} {3rd}\ ed.\
  (\bibinfo  {publisher} {{Wiley}},\ \bibinfo {address} {{New York}},\ \bibinfo
  {year} {1998})\BibitemShut {NoStop}%
\bibitem [{\citenamefont {Liu}(2017)}]{LIU_Phys.Rev.Lett._generalized_2017}%
  \BibitemOpen
  \bibfield  {author} {\bibinfo {author} {\bibfnamefont {W.}~\bibnamefont
  {Liu}},\ }\bibfield  {title} {\bibinfo {title} {Generalized magnetic
  mirrors},\ }\href@noop {} {\bibfield  {journal} {\bibinfo  {journal} {Phys.
  Rev. Lett.}\ }\textbf {\bibinfo {volume} {119}},\ \bibinfo {pages} {123902}
  (\bibinfo {year} {2017})}\BibitemShut {NoStop}%
\bibitem [{\citenamefont {Chen}\ \emph {et~al.}(2019)\citenamefont {Chen},
  \citenamefont {Chen},\ and\ \citenamefont {Liu}}]{CHEN_2019__Singularities}%
  \BibitemOpen
  \bibfield  {author} {\bibinfo {author} {\bibfnamefont {W.}~\bibnamefont
  {Chen}}, \bibinfo {author} {\bibfnamefont {Y.}~\bibnamefont {Chen}},\ and\
  \bibinfo {author} {\bibfnamefont {W.}~\bibnamefont {Liu}},\ }\bibfield
  {title} {\bibinfo {title} {Singularities and {{P}}oincar\'e indices of
  electromagnetic multipoles},\ }\href@noop {} {\bibfield  {journal} {\bibinfo
  {journal} {Phys. Rev. Lett.}\ }\textbf {\bibinfo {volume} {122}},\ \bibinfo
  {pages} {153907} (\bibinfo {year} {2019})}\BibitemShut {NoStop}%
\bibitem [{\citenamefont {Chen}\ \emph {et~al.}(2020)\citenamefont {Chen},
  \citenamefont {Chen},\ and\ \citenamefont
  {Liu}}]{CHEN_2019_ArXiv190409910Math-PhPhysicsphysics_Linea}%
  \BibitemOpen
  \bibfield  {author} {\bibinfo {author} {\bibfnamefont {W.}~\bibnamefont
  {Chen}}, \bibinfo {author} {\bibfnamefont {Y.}~\bibnamefont {Chen}},\ and\
  \bibinfo {author} {\bibfnamefont {W.}~\bibnamefont {Liu}},\ }\bibfield
  {title} {\bibinfo {title} {Line {{Singularities}} and {{Hopf Indices}} of
  {{Electromagnetic Multipoles}}},\ }\href@noop {} {\bibfield  {journal}
  {\bibinfo  {journal} {Laser Photonics Rev.}\ }\textbf {\bibinfo {volume}
  {14}},\ \bibinfo {pages} {2000049} (\bibinfo {year} {2020})}\BibitemShut
  {NoStop}%
\bibitem [{\citenamefont {{Fernandez-Corbaton}}\ \emph
  {et~al.}(2013)\citenamefont {{Fernandez-Corbaton}}, \citenamefont
  {{Zambrana-Puyalto}}, \citenamefont {Tischler}, \citenamefont {Vidal},
  \citenamefont {Juan},\ and\ \citenamefont
  {{Molina-Terriza}}}]{FERNANDEZ-CORBATON_2013_Phys.Rev.Lett._Electromagnetica}%
  \BibitemOpen
  \bibfield  {author} {\bibinfo {author} {\bibfnamefont {I.}~\bibnamefont
  {{Fernandez-Corbaton}}}, \bibinfo {author} {\bibfnamefont {X.}~\bibnamefont
  {{Zambrana-Puyalto}}}, \bibinfo {author} {\bibfnamefont {N.}~\bibnamefont
  {Tischler}}, \bibinfo {author} {\bibfnamefont {X.}~\bibnamefont {Vidal}},
  \bibinfo {author} {\bibfnamefont {M.~L.}\ \bibnamefont {Juan}},\ and\
  \bibinfo {author} {\bibfnamefont {G.}~\bibnamefont {{Molina-Terriza}}},\
  }\bibfield  {title} {\bibinfo {title} {Electromagnetic {{Duality Symmetry}}
  and {{Helicity Conservation}} for the {{Macroscopic Maxwell}}'s
  {{Equations}}},\ }\href@noop {} {\bibfield  {journal} {\bibinfo  {journal}
  {Phys. Rev. Lett.}\ }\textbf {\bibinfo {volume} {111}},\ \bibinfo {pages}
  {060401} (\bibinfo {year} {2013})}\BibitemShut {NoStop}%
\bibitem [{\citenamefont {Yang}\ \emph {et~al.}(2020)\citenamefont {Yang},
  \citenamefont {Chen}, \citenamefont {Chen},\ and\ \citenamefont
  {Liu}}]{YANG_2020_ACSPhotonics_Electromagnetic}%
  \BibitemOpen
  \bibfield  {author} {\bibinfo {author} {\bibfnamefont {Q.}~\bibnamefont
  {Yang}}, \bibinfo {author} {\bibfnamefont {W.}~\bibnamefont {Chen}}, \bibinfo
  {author} {\bibfnamefont {Y.}~\bibnamefont {Chen}},\ and\ \bibinfo {author}
  {\bibfnamefont {W.}~\bibnamefont {Liu}},\ }\bibfield  {title} {\bibinfo
  {title} {Electromagnetic {{Duality Protected Scattering Properties}} of
  {{Nonmagnetic Particles}}},\ }\href@noop {} {\bibfield  {journal} {\bibinfo
  {journal} {ACS Photonics}\ }\textbf {\bibinfo {volume} {7}},\ \bibinfo
  {pages} {1830} (\bibinfo {year} {2020})}\BibitemShut {NoStop}%
\bibitem [{\citenamefont {Lalanne}\ \emph {et~al.}(2018)\citenamefont
  {Lalanne}, \citenamefont {Yan}, \citenamefont {Vynck}, \citenamefont
  {Sauvan},\ and\ \citenamefont {Hugonin}}]{LALANNE__LaserPhotonicsRev._Light}%
  \BibitemOpen
  \bibfield  {author} {\bibinfo {author} {\bibfnamefont {P.}~\bibnamefont
  {Lalanne}}, \bibinfo {author} {\bibfnamefont {W.}~\bibnamefont {Yan}},
  \bibinfo {author} {\bibfnamefont {K.}~\bibnamefont {Vynck}}, \bibinfo
  {author} {\bibfnamefont {C.}~\bibnamefont {Sauvan}},\ and\ \bibinfo {author}
  {\bibfnamefont {J.-P.}\ \bibnamefont {Hugonin}},\ }\bibfield  {title}
  {\bibinfo {title} {Light {{Interaction}} with {{Photonic}} and {{Plasmonic
  Resonances}}},\ }\href@noop {} {\bibfield  {journal} {\bibinfo  {journal}
  {Laser Photonics Rev.}\ }\textbf {\bibinfo {volume} {12}},\ \bibinfo {pages}
  {1700113} (\bibinfo {year} {2018})}\BibitemShut {NoStop}%
\bibitem [{\citenamefont {Gras}\ \emph {et~al.}(2020)\citenamefont {Gras},
  \citenamefont {Lalanne},\ and\ \citenamefont
  {Durufl{\'e}}}]{GRAS_2020_J.Opt.Soc.Am.A_Nonuniqueness}%
  \BibitemOpen
  \bibfield  {author} {\bibinfo {author} {\bibfnamefont {A.}~\bibnamefont
  {Gras}}, \bibinfo {author} {\bibfnamefont {P.}~\bibnamefont {Lalanne}},\ and\
  \bibinfo {author} {\bibfnamefont {M.}~\bibnamefont {Durufl{\'e}}},\
  }\bibfield  {title} {\bibinfo {title} {Nonuniqueness of the quasinormal mode
  expansion of electromagnetic {{Lorentz}} dispersive materials},\ }\href
  {https://doi.org/10.1364/JOSAA.394206} {\bibfield  {journal} {\bibinfo
  {journal} {J. Opt. Soc. Am. A}\ }\textbf {\bibinfo {volume} {37}},\ \bibinfo
  {pages} {1219} (\bibinfo {year} {2020})}\BibitemShut {NoStop}%
\bibitem [{\citenamefont {Caloz}\ \emph {et~al.}(2018)\citenamefont {Caloz},
  \citenamefont {Al{\`u}}, \citenamefont {Tretyakov}, \citenamefont {Sounas},
  \citenamefont {Achouri},\ and\ \citenamefont
  {{Deck-L{\'e}ger}}}]{CALOZ_2018_Phys.Rev.Applied_Electromagnetic}%
  \BibitemOpen
  \bibfield  {author} {\bibinfo {author} {\bibfnamefont {C.}~\bibnamefont
  {Caloz}}, \bibinfo {author} {\bibfnamefont {A.}~\bibnamefont {Al{\`u}}},
  \bibinfo {author} {\bibfnamefont {S.}~\bibnamefont {Tretyakov}}, \bibinfo
  {author} {\bibfnamefont {D.}~\bibnamefont {Sounas}}, \bibinfo {author}
  {\bibfnamefont {K.}~\bibnamefont {Achouri}},\ and\ \bibinfo {author}
  {\bibfnamefont {Z.-L.}\ \bibnamefont {{Deck-L{\'e}ger}}},\ }\bibfield
  {title} {\bibinfo {title} {Electromagnetic {{Nonreciprocity}}},\ }\href
  {https://doi.org/10.1103/PhysRevApplied.10.047001} {\bibfield  {journal}
  {\bibinfo  {journal} {Phys. Rev. Applied}\ }\textbf {\bibinfo {volume}
  {10}},\ \bibinfo {pages} {047001} (\bibinfo {year} {2018})}\BibitemShut
  {NoStop}%
\bibitem [{\citenamefont
  {Pancharatnam}(1956)}]{PANCHARATNAM_1956_ProcIndianAcadSci_Generalized}%
  \BibitemOpen
  \bibfield  {author} {\bibinfo {author} {\bibfnamefont {S.}~\bibnamefont
  {Pancharatnam}},\ }\bibfield  {title} {\bibinfo {title} {Generalized theory
  of interference, and its applications},\ }\href@noop {} {\bibfield  {journal}
  {\bibinfo  {journal} {Proc. Indian. Acad. Sci.}\ }\textbf {\bibinfo {volume}
  {44}},\ \bibinfo {pages} {247} (\bibinfo {year} {1956})}\BibitemShut
  {NoStop}%
\bibitem [{\citenamefont {Berry}(1984)}]{BERRY_1984_Proc.R.Soc.A_Quantal}%
  \BibitemOpen
  \bibfield  {author} {\bibinfo {author} {\bibfnamefont {M.~V.}\ \bibnamefont
  {Berry}},\ }\bibfield  {title} {\bibinfo {title} {Quantal {{Phase Factors
  Accompanying Adiabatic Changes}}},\ }\href@noop {} {\bibfield  {journal}
  {\bibinfo  {journal} {Proc. R. Soc. A}\ }\textbf {\bibinfo {volume} {392}},\
  \bibinfo {pages} {45} (\bibinfo {year} {1984})}\BibitemShut {NoStop}%
\bibitem [{\citenamefont {Berry}(1987)}]{berry_adiabatic_1987}%
  \BibitemOpen
  \bibfield  {author} {\bibinfo {author} {\bibfnamefont {M.~V.}\ \bibnamefont
  {Berry}},\ }\bibfield  {title} {\bibinfo {title} {The {{Adiabatic Phase}} and
  {{Pancharatnam Phase}} for {{Polarized}}-{{Light}}},\ }\href@noop {}
  {\bibfield  {journal} {\bibinfo  {journal} {J. Mod. Opt.}\ }\textbf {\bibinfo
  {volume} {34}},\ \bibinfo {pages} {1401} (\bibinfo {year}
  {1987})}\BibitemShut {NoStop}%
\bibitem [{\citenamefont
  {Berry}(2024)}]{BERRY_2024_Opt.PhotonicsNews_GeometricPhase}%
  \BibitemOpen
  \bibfield  {author} {\bibinfo {author} {\bibfnamefont {M.}~\bibnamefont
  {Berry}},\ }\bibfield  {title} {\bibinfo {title} {A {{Geometric-Phase
  Timeline}}},\ }\href@noop {} {\bibfield  {journal} {\bibinfo  {journal}
  {Optics \& Photonics News}\ }\textbf {\bibinfo {volume} {3}},\ \bibinfo
  {pages} {42} (\bibinfo {year} {2024})}\BibitemShut {NoStop}%
\bibitem [{\citenamefont {Yariv}\ and\ \citenamefont
  {Yeh}(2006)}]{YARIV_2006__Photonics}%
  \BibitemOpen
  \bibfield  {author} {\bibinfo {author} {\bibfnamefont {A.}~\bibnamefont
  {Yariv}}\ and\ \bibinfo {author} {\bibfnamefont {P.}~\bibnamefont {Yeh}},\
  }\href@noop {} {\emph {\bibinfo {title} {Photonics: {{Optical Electronics}}
  in {{Modern Communications}}}}},\ \bibinfo {edition} {6th}\ ed.\ (\bibinfo
  {publisher} {{Oxford University Press}},\ \bibinfo {address} {{New York}},\
  \bibinfo {year} {2006})\BibitemShut {NoStop}%
\bibitem [{\citenamefont {Chen}\ \emph {et~al.}(2021)\citenamefont {Chen},
  \citenamefont {Yang}, \citenamefont {Chen},\ and\ \citenamefont
  {Liu}}]{CHEN_Phys.Rev.Lett._Extremize}%
  \BibitemOpen
  \bibfield  {author} {\bibinfo {author} {\bibfnamefont {W.}~\bibnamefont
  {Chen}}, \bibinfo {author} {\bibfnamefont {Q.}~\bibnamefont {Yang}}, \bibinfo
  {author} {\bibfnamefont {Y.}~\bibnamefont {Chen}},\ and\ \bibinfo {author}
  {\bibfnamefont {W.}~\bibnamefont {Liu}},\ }\bibfield  {title} {\bibinfo
  {title} {Extremize {{Optical Chiralities}} through {{Polarization
  Singularities}}},\ }\href {https://doi.org/10.1103/PhysRevLett.126.253901}
  {\bibfield  {journal} {\bibinfo  {journal} {Phys. Rev. Lett.}\ }\textbf
  {\bibinfo {volume} {126}},\ \bibinfo {pages} {253901} (\bibinfo {year}
  {2021})}\BibitemShut {NoStop}%
\bibitem [{\citenamefont {Ramachandran}\ and\ \citenamefont
  {Ramaseshan}(1961)}]{RAMACHANDRAN_1961}%
  \BibitemOpen
  \bibfield  {author} {\bibinfo {author} {\bibfnamefont {G.~N.}\ \bibnamefont
  {Ramachandran}}\ and\ \bibinfo {author} {\bibfnamefont {S.}~\bibnamefont
  {Ramaseshan}},\ }\bibfield  {title} {\bibinfo {title} {Crystal {{Optics}}},\
  }in\ \href {https://doi.org/10.1007/978-3-642-45959-7_1} {\emph {\bibinfo
  {booktitle} {Kristalloptik {$\cdot$} {{Beugung}}/{{Crystal Optics}} {$\cdot$}
  {{Diffraction}}}}},\ \bibinfo {series and number} {Handbuch Der {{Physik}} /
  {{Encyclopedia}} of {{Physics}}},\ \bibinfo {editor} {edited by\ \bibinfo
  {editor} {\bibfnamefont {S.}~\bibnamefont {Fl{\"u}gge}}}\ (\bibinfo
  {publisher} {Springer},\ \bibinfo {address} {Berlin, Heidelberg},\ \bibinfo
  {year} {1961})\ pp.\ \bibinfo {pages} {1--217}\BibitemShut {NoStop}%
\bibitem [{\citenamefont {Pancharatnam}(1975)}]{PANCHARATNAM_1975__Collecteda}%
  \BibitemOpen
  \bibfield  {author} {\bibinfo {author} {\bibfnamefont {S.}~\bibnamefont
  {Pancharatnam}},\ }\href@noop {} {\emph {\bibinfo {title} {Collected Works of
  {{S}}. {{Pancharatnam}}}}}\ (\bibinfo  {publisher} {{Oxford University Press
  for the Raman Research Institute}},\ \bibinfo {address} {{London}},\ \bibinfo
  {year} {1975})\BibitemShut {NoStop}%
\bibitem [{Sup()}]{Supplemental_Material}%
  \BibitemOpen
  \href@noop {} {\  {\bibinfo {volume} {See Supplemental Material that
  includes the following seven sections:
  (\textbf{\uppercase\expandafter{\romannumeral1}}). Relative amplitudes and
  phases among QNM expansion coefficients;
  (\textbf{\uppercase\expandafter{\romannumeral2}}). Complex eigenfrequencies
  and near fields of QNMs, and their scattered field intensity and polarization
  distributions on the momentum sphere;
  (\textbf{\uppercase\expandafter{\romannumeral3}}). Normalized Jones vectors
  and Stokes parameters for polarizations of QNM radiations opposite to the
  incident direction; (\textbf{\uppercase\expandafter{\romannumeral4}}).
  Selective QNM excitations; (\textbf{\uppercase\expandafter{\romannumeral5}}).
  Non-perpendicular bi-cylinder scattering;
  (\textbf{\uppercase\expandafter{\romannumeral6}}). Twisted bi-SRR scattering;
  (\textbf{\uppercase\expandafter{\romannumeral7}}). Gain particle scattering.
  Supplemental Material includes
  Refs.~\cite{WEN_2024_Laser&PhotonicsReviews_Momentum,BERRY_2024_Opt.PhotonicsNews_GeometricPhase,Bohren1983_book,CHEN_Phys.Rev.Lett._Extremize,YARIV_2006__Photonics,Johnson1972_PRB}}}}\BibitemShut
  {NoStop}%
\bibitem [{\citenamefont {Wen}\ \emph {et~al.}(2024)\citenamefont {Wen},
  \citenamefont {Zhang}, \citenamefont {Qin}, \citenamefont {Zhu},\ and\
  \citenamefont {Liu}}]{WEN_2024_Laser&PhotonicsReviews_Momentum}%
  \BibitemOpen
  \bibfield  {author} {\bibinfo {author} {\bibfnamefont {C.}~\bibnamefont
  {Wen}}, \bibinfo {author} {\bibfnamefont {J.}~\bibnamefont {Zhang}}, \bibinfo
  {author} {\bibfnamefont {S.}~\bibnamefont {Qin}}, \bibinfo {author}
  {\bibfnamefont {Z.}~\bibnamefont {Zhu}},\ and\ \bibinfo {author}
  {\bibfnamefont {W.}~\bibnamefont {Liu}},\ }\bibfield  {title} {\bibinfo
  {title} {Momentum-{{Space Scattering Extremizations}}},\ }\href
  {https://doi.org/10.1002/lpor.202300454} {\bibfield  {journal} {\bibinfo
  {journal} {Laser \& Photonics Reviews}\ }\textbf {\bibinfo {volume} {18}},\
  \bibinfo {pages} {2300454} (\bibinfo {year} {2024})}\BibitemShut {NoStop}%
\bibitem [{\citenamefont {Johnson}\ and\ \citenamefont
  {Christy}(1972)}]{Johnson1972_PRB}%
  \BibitemOpen
  \bibfield  {author} {\bibinfo {author} {\bibfnamefont {P.~B.}\ \bibnamefont
  {Johnson}}\ and\ \bibinfo {author} {\bibfnamefont {R.~W.}\ \bibnamefont
  {Christy}},\ }\bibfield  {title} {\bibinfo {title} {Optical constants of the
  noble metals},\ }\href@noop {} {\bibfield  {journal} {\bibinfo  {journal}
  {Phys. Rev. B}\ }\textbf {\bibinfo {volume} {6}},\ \bibinfo {pages} {4370}
  (\bibinfo {year} {1972})}\BibitemShut {NoStop}%
\bibitem [{\citenamefont {Li}\ \emph {et~al.}(2017)\citenamefont {Li},
  \citenamefont {Zhang},\ and\ \citenamefont
  {Zentgraf}}]{LI_Nat.Rev.Mater._nonlinear_2017}%
  \BibitemOpen
  \bibfield  {author} {\bibinfo {author} {\bibfnamefont {G.}~\bibnamefont
  {Li}}, \bibinfo {author} {\bibfnamefont {S.}~\bibnamefont {Zhang}},\ and\
  \bibinfo {author} {\bibfnamefont {T.}~\bibnamefont {Zentgraf}},\ }\bibfield
  {title} {\bibinfo {title} {Nonlinear photonic metasurfaces},\ }\href@noop {}
  {\bibfield  {journal} {\bibinfo  {journal} {Nat. Rev. Mater.}\ }\textbf
  {\bibinfo {volume} {2}},\ \bibinfo {pages} {17010} (\bibinfo {year}
  {2017})}\BibitemShut {NoStop}%
\bibitem [{\citenamefont {Cohen}\ \emph {et~al.}(2019)\citenamefont {Cohen},
  \citenamefont {Larocque}, \citenamefont {Bouchard}, \citenamefont
  {Nejadsattari}, \citenamefont {Gefen},\ and\ \citenamefont
  {Karimi}}]{COHEN_2019_NatRevPhys_Geometric}%
  \BibitemOpen
  \bibfield  {author} {\bibinfo {author} {\bibfnamefont {E.}~\bibnamefont
  {Cohen}}, \bibinfo {author} {\bibfnamefont {H.}~\bibnamefont {Larocque}},
  \bibinfo {author} {\bibfnamefont {F.}~\bibnamefont {Bouchard}}, \bibinfo
  {author} {\bibfnamefont {F.}~\bibnamefont {Nejadsattari}}, \bibinfo {author}
  {\bibfnamefont {Y.}~\bibnamefont {Gefen}},\ and\ \bibinfo {author}
  {\bibfnamefont {E.}~\bibnamefont {Karimi}},\ }\bibfield  {title} {\bibinfo
  {title} {Geometric phase from {{Aharonov}}--{{Bohm}} to
  {{Pancharatnam}}--{{Berry}} and beyond},\ }\href
  {https://doi.org/10.1038/s42254-019-0071-1} {\bibfield  {journal} {\bibinfo
  {journal} {Nat. Rev. Phys.}\ }\textbf {\bibinfo {volume} {1}},\ \bibinfo
  {pages} {437} (\bibinfo {year} {2019})}\BibitemShut {NoStop}%
\bibitem [{\citenamefont {Jisha}\ \emph {et~al.}(2021)\citenamefont {Jisha},
  \citenamefont {Nolte},\ and\ \citenamefont
  {Alberucci}}]{JISHA_2021_LaserPhotonicsRev._Geometric}%
  \BibitemOpen
  \bibfield  {author} {\bibinfo {author} {\bibfnamefont {C.~P.}\ \bibnamefont
  {Jisha}}, \bibinfo {author} {\bibfnamefont {S.}~\bibnamefont {Nolte}},\ and\
  \bibinfo {author} {\bibfnamefont {A.}~\bibnamefont {Alberucci}},\ }\bibfield
  {title} {\bibinfo {title} {Geometric {{Phase}} in {{Optics}}: {{From
  Wavefront Manipulation}} to {{Waveguiding}}},\ }\href
  {https://doi.org/10.1002/lpor.202100003} {\bibfield  {journal} {\bibinfo
  {journal} {Laser \& Photonics Reviews}\ }\textbf {\bibinfo {volume} {15}},\
  \bibinfo {pages} {2100003} (\bibinfo {year} {2021})}\BibitemShut {NoStop}%
\bibitem [{\citenamefont {Guo}\ \emph {et~al.}(2022)\citenamefont {Guo},
  \citenamefont {Pu}, \citenamefont {Zhang}, \citenamefont {Xu}, \citenamefont
  {Li}, \citenamefont {Ma},\ and\ \citenamefont {Luo}}]{GUO_2022_PI_Classical}%
  \BibitemOpen
  \bibfield  {author} {\bibinfo {author} {\bibfnamefont {Y.}~\bibnamefont
  {Guo}}, \bibinfo {author} {\bibfnamefont {M.}~\bibnamefont {Pu}}, \bibinfo
  {author} {\bibfnamefont {F.}~\bibnamefont {Zhang}}, \bibinfo {author}
  {\bibfnamefont {M.}~\bibnamefont {Xu}}, \bibinfo {author} {\bibfnamefont
  {X.}~\bibnamefont {Li}}, \bibinfo {author} {\bibfnamefont {X.}~\bibnamefont
  {Ma}},\ and\ \bibinfo {author} {\bibfnamefont {X.}~\bibnamefont {Luo}},\
  }\bibfield  {title} {\bibinfo {title} {Classical and generalized geometric
  phase in electromagnetic metasurfaces},\ }\href
  {https://doi.org/10.3788/PI.2022.R03} {\bibfield  {journal} {\bibinfo
  {journal} {Photonics Insights}\ }\textbf {\bibinfo {volume} {1}},\ \bibinfo
  {pages} {R03} (\bibinfo {year} {2022})}\BibitemShut {NoStop}%
\bibitem [{\citenamefont {Gbur}(2016)}]{GBUR_2016__Singular}%
  \BibitemOpen
  \bibfield  {author} {\bibinfo {author} {\bibfnamefont {G.~J.}\ \bibnamefont
  {Gbur}},\ }\href@noop {} {\emph {\bibinfo {title} {Singular {{Optics}}}}}\
  (\bibinfo  {publisher} {{CRC Press Inc}},\ \bibinfo {address} {Boca Raton},\
  \bibinfo {year} {2016})\BibitemShut {NoStop}%
\end{thebibliography}

%


\onecolumngrid
\clearpage


{\centering
  \noindent\textbf{\large{Supplemental Material for:}}
\\\bigskip
\noindent\textbf{\large{Geometric Phase-Driven Scattering Evolutions}}
\\\bigskip
\onecolumngrid

Pengxiang Wang$^{1}$,  Yuntian Chen$^{1,2,\dag}$, and Wei Liu$^{3,4,\ddagger}$

\small{$^1$ \emph{School of Optical and Electronic Information, Huazhong University of Science and Technology, Wuhan, Hubei 430074, P. R. China}}\\
\small{$^2$ \emph{Wuhan National Laboratory for Optoelectronics, Huazhong University of Science and Technology, Wuhan, Hubei 430074, P. R. China}}\\
\small{$^3$ \emph{College for Advanced Interdisciplinary Studies, National University of Defense
Technology, Changsha, Hunan 410073, P. R. China}}\\
\small{$^4$ \emph{Nanhu Laser Laboratory and Hunan Provincial Key Laboratory of Novel Nano-Optoelectronic Information Materials and
Devices, National University of Defense Technology, Changsha 410073, P. R. China}}\par
}

The Supplemental Material includes the following seven sections: (\textbf{\uppercase\expandafter{\romannumeral1}}). Relative amplitudes and phases among QNM expansion coefficients;
(\textbf{\uppercase\expandafter{\romannumeral2}}). Complex eigenfrequencies and near fields of QNMs, and their scattered field intensity and polarization distributions on the momentum sphere;
(\textbf{\uppercase\expandafter{\romannumeral3}}). Normalized Jones vectors and Stokes parameters for polarizations of QNM radiations opposite to the incident direction; (\textbf{\uppercase\expandafter{\romannumeral4}}). Selective QNM excitations;
(\textbf{\uppercase\expandafter{\romannumeral5}}). Non-perpendicular bi-cylinder scattering;
(\textbf{\uppercase\expandafter{\romannumeral6}}). Twisted bi-SRR scattering;
(\textbf{\uppercase\expandafter{\romannumeral7}}). Gain particle scattering.\\

\setcounter{equation}{0}
\setcounter{figure}{0}
\newcounter{sfigure}
\setcounter{sfigure}{1}
\setcounter{table}{0}
\renewcommand{\theequation}{S\arabic{equation}}

 \renewcommand\thefigure{S{\arabic{figure}}}
\renewcommand{\thesection}{S\arabic{section}}

\section{(\textbf{\uppercase\expandafter{\romannumeral1}}). Relative amplitudes and phases among QNM expansion coefficients}
In this study, we expand the scattered fields into radiations of QNMs rather than electromagnetic multipoles (spherical harmonics). In this paper we focus on far-field scattering properties, and the expansion can be expressed as:
\begin{equation}
\label{expansion-sm}
\mathbf{E}_{\rm{sca}}(\mathbf{{\hat{r}}})=\sum_m \alpha_{m} {\mathbf{E}}_{m}(\mathbf{{\hat{r}}}).
\end{equation}
After the QNMs are calculated, all scattering manipulations rely on controlling the relative amplitudes and phases among  $\alpha_{m}$. As we have revealed, when the incident field excites modes A and B, the amplitudes satisfy: 
\begin{equation}
\label{amplitude-sm}
|\alpha_A| \propto \cos\left(\frac{1}{2}\arc{PA}\right),~ |\alpha_B| \propto \cos\left(\frac{1}{2}\arc{PB}\right)
\end{equation}
and the relative phase can be expressed as:
\begin{equation}
\label{phase-sm}
\mathrm{Arg}(\alpha_B)-\mathrm{Arg}(\alpha_A)=\varphi_g.
\end{equation}
Based on Eqs.~(\ref{amplitude-sm}) and (\ref{phase-sm}), Eq.~(\ref{expansion-sm}) can be reformulated as [Eq. (4) in the main Letter]:
\begin{equation}
\label{expansion2-sm}
\mathbf{E}_{\rm{sca}}(\mathbf{{\hat{r}}}) \propto \cos\left(\frac{1}{2}\arc{PA}\right)~{\mathbf{E}}_{\rm{A}}(\mathbf{{\hat{r}}})+\cos\left(\frac{1}{2}\arc{PB}\right)~{\mathbf{E}}_{\rm{B}}(\mathbf{{\hat{r}}}) \exp({i\varphi_g}). 
\end{equation}

It is worth mentioning that though Eq.~(\ref{expansion2-sm}) is the simplest expression for the QNM expansion, it is not the unique one. This is due to the fact that ${\mathbf{E}}_{\rm{A,B}}(\mathbf{{\hat{r}}})$ also have their own amplitudes and phases. It is clear that Eq.~(\ref{amplitude-sm}) measures only the polarization discrepancy between QNM radiation (opposite to the incident direction) and incident field,  while containing no information about the QNM radiation intensity.  It is recently revealed that the amplitudes of expansion coefficients are related to both QNM radiation polarizations and intensities~\cite{WEN_2024_Laser&PhotonicsReviews_Momentum}. Nevertheless, Eq.~(\ref{expansion2-sm}) can still accommodate the scenarios of identical QNM radiation polarization [\textit{e.g.} $\cos(\frac{1}{2}\arc{PA})$ is fixed] while different radiation intensities by adopting accordingly the QNM field [\textit{e.g.} ${\mathbf{E}}_{\rm{A}}(\mathbf{{\hat{r}}})$] of a proper amplitude. Similarly, though the phase for each QNM can be arbitrarily chosen due to gauge freedom of electromagnetism~\cite{BERRY_2024_Opt.PhotonicsNews_GeometricPhase},  the scattered field can always be expressed by Eq.~(\ref{expansion2-sm}) by assigning accordingly to ${\mathbf{E}}_{\rm{A,B}}(\mathbf{{\hat{r}}})$ proper phases.  The non-geometric relative phase between ${\mathbf{E}}_{\rm{A,B}}(\mathbf{{\hat{r}}})$ can be fixed by the optical theorem, which imposes a stringent constraint on both the phase and amplitude of the forward scattering $\mathbf{E}_{\rm{sca}}(\mathbf{\hat{r}}_{\rm{inc}})$~\cite{Bohren1983_book}. Alternatively, it can be quickly decided through a single fitting of Eq.~(\ref{expansion2-sm}) with numerically calculated results. This non-geometric phase contrast obtained can then be used for any incident polarizations, though generally it is dependent on the incident direction, along which the optical theorem imposes its constraint.  

Another factor that affects the relative amplitudes and phases among $\alpha_{m}$ is the frequency detuning, \textit{ e.g.} incident frequency deviates from the central frequency of the QNM. Such a deviation would induce both amplitude and phase shift for $\alpha_{m}$, depending on the Lorentzian line shape of the corresponding QNM. Those extra phase and amplitude shifts can be both absorbed into   ${\mathbf{E}}_{\rm{A,B}}(\mathbf{{\hat{r}}})$, leaving Eq.~(\ref{expansion2-sm}) still valid.

With the expression for the scattered field [Eq.~(\ref{expansion2-sm})], the scattering intensity is:
\begin{equation}
\label{expansion-simplified-intensity2}
\begin{split}
I_{\rm{sca}} = |\mathbf{E}_{\rm{sca}}|^2 \propto \cos^2\left(\frac{1}{2}\arc{PA}\right)I_{\rm{A}}+\cos^2\left(\frac{1}{2}\arc{PB}\right)I_{\rm{B}}\\+
2\cos(\frac{1}{2}\arc{PA})\cos\left(\frac{1}{2}\arc{PB}\right)\sqrt{I_{\rm{A}}I_{\rm{B}}}|\mathbb{J}_{\rm{A}}\mathbb{J}_{\rm{B}}^{\dagger}|\cos(\varphi_g),  
\end{split}
\end{equation}
where $I_{\rm{A,B}}$ denotes QNM radiation intensity: $I_{\rm{A,B}}=|{\mathbf{E}}_{\rm{A,B}}(\mathbf{\hat{r}})|^2$. When the incident polarization is orthogonal to that of mode A along $-\mathbf{\hat{r}}_{\rm{i}}$ ($\mathbf{J_{\rm{i}}}\mathbf{J}^{\dagger}_{\rm{A}}=0$), P and A are antipodal points ($\arc{PA}=\pi$) and thus mode A would not be excited [$\cos(\frac{1}{2}\arc{PA})=0$], being consistent with the special scenario of single-mode excitations~\cite{CHEN_Phys.Rev.Lett._Extremize,WEN_2024_Laser&PhotonicsReviews_Momentum} [see also Section (\textbf{\uppercase\expandafter{\romannumeral4}}) for specific demonstrations].

\section{(\textbf{\uppercase\expandafter{\romannumeral2}}). Complex eigenfrequencies and near fields of QNMs, and their scattered field intensity and polarization distributions on the momentum sphere}

\begin{figure}[htbp]
\centerline{\includegraphics[width=12cm]{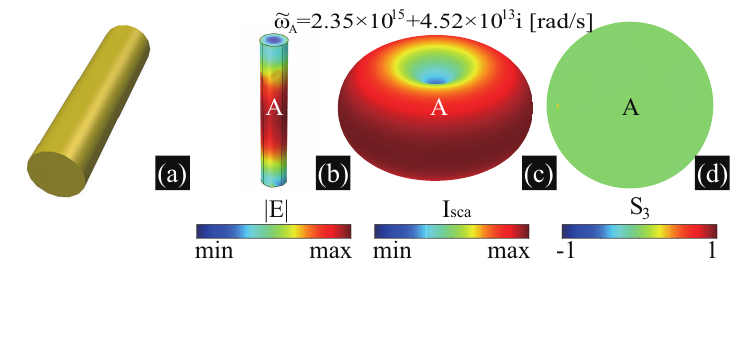}} \caption{\small (a) An individual cylinder supports a QNM of electric dipolar nature, with its complex eigenfrequency specified.  (b) Near fields (in terms of total electric field amplitude $\mathbf{|E|}$) of the QNM. Far-field scattering intensity (c) and polarization in terms of $S_3$ (d) distributions on the momentum sphere. }
\label{figure1_SP}
\end{figure}

\begin{figure}[htbp]
\centerline{\includegraphics[width=12cm]{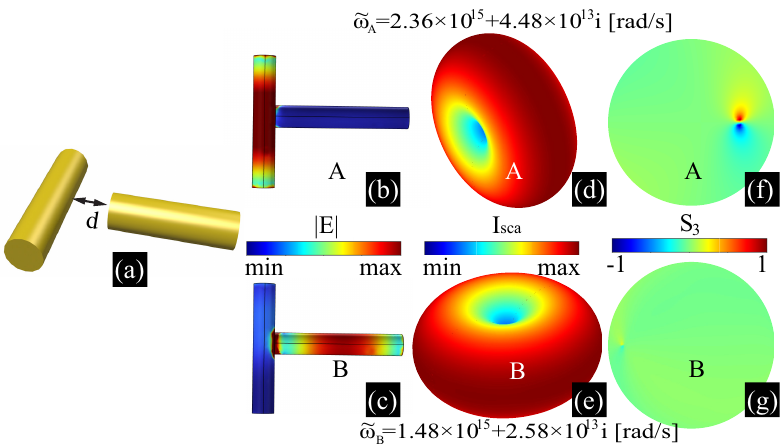}} \caption{\small \small (a) A pair of perpendicular cylinders supports two QNMs with their complex eigenfrequencies specified (when $d=0$).  (b) \& (c) Near fields (in terms of total electric field amplitude $\mathbf{|E|}$) of the QNMs. Far-field scattering intensity [(d) \& (e)] and polarization in terms of $S_3$ [(f) \& (g)] distributions on the momentum sphere.}
\label{figure2_SP}
\end{figure}

\begin{figure}[htbp]
\centerline{\includegraphics[width=12cm]{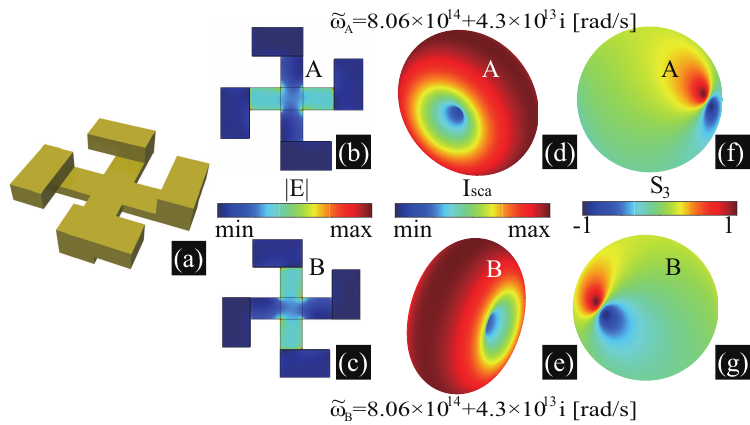}} \caption{\small (a) A gold particle with $4$-fold rotation symmetry supports two QNMs with their complex eigenfrequencies specified.  (b) \& (c) Near fields (in terms of total electric field amplitude $\mathbf{|E|}$) of the QNMs. Far-field scattering intensity [(d) \& (e)] and polarization in terms of $S_3$ [(f) \& (g)] distributions on the momentum sphere.}
\label{figure3_SP}
\end{figure}

\begin{figure}[htbp]
\centerline{\includegraphics[width=12cm]{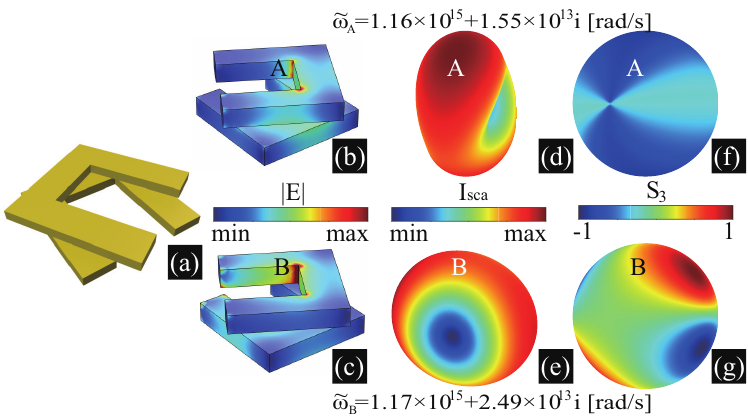}} \caption{\small (a) A pair of twisted split ring resonators supports two QNMs with their complex eigenfrequencies specified.  (b) \& (c) Near fields (in terms of total electric field amplitude $\mathbf{|E|}$) of the QNMs. Far-field scattering intensity [(d) \& (e)] and polarization in terms of $S_3$ [(f) \& (g)] distributions on the momentum sphere.}
\label{figure4_SP}
\end{figure}

\begin{figure}[htbp]
\centerline{\includegraphics[width=12cm]{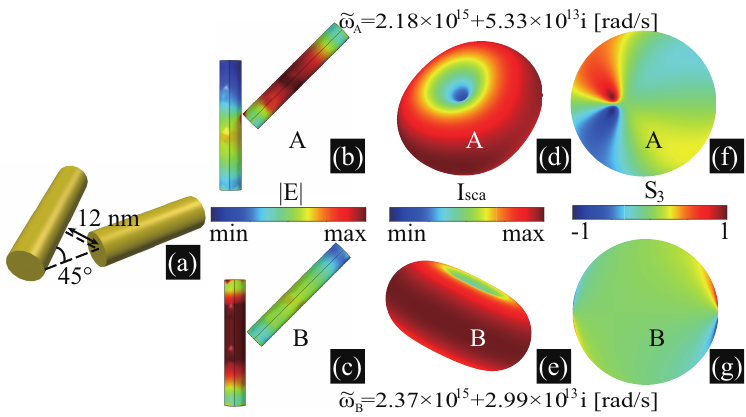}} \caption{\small (a) A pair of non-perpendicular cylinders (one twisted by $45^{\circ}$ with respect to the other) supports two QNMs with their complex eigenfrequencies specified.  (b) \& (c) Near fields (in terms of total electric field amplitude $\mathbf{|E|}$) of the QNMs. Far-field scattering intensity [(d) \& (e)] and polarization in terms of $S_3$ [(f) \& (g)] distributions on the momentum sphere.}
\label{figure5_SP}
\end{figure}

In Figs. \ref{figure1_SP}-\ref{figure5_SP} we have summarized the QNM information (complex eigenfrequencies, near fields, scattered far field intensity and polarization distributions on the momentum sphere) for all structures studied in both the main Letter and this Supplemental Material.

\section{(\textbf{\uppercase\expandafter{\romannumeral3}}). Normalized Jones vectors and Stokes parameters for polarizations of QNM radiations opposite to the incident direction}

In Table \ref{table} we have summarized the normalized Jones vectors and Stokes parameters ($S_1$, $S_2$, $S_3$) for polarizations of QNMs (modes A and B supported by different structures studied in both the main Letter and this Supplemental Material) radiations opposite to the incident direction. 

\begin{table}[htp]
\centering 

\begin{tabular}{|c|c|c|}
\hline & Stokes parameters  & Normalized Jones vector \\
\hline Fig.1(e) & A: :[1,0,0] & A: $(1,0)$ \\
& B: $[-1,0,0]$ & B: $(0,1)$ \\
\hline Fig.1(f) & A: $[1,0,0]$ & A: $(1,0)$ \\
& B: $[-0.92,0.316,-0.236]$ & B: $(0.2,0.785-0.586 i)$ \\
\hline Figs. 2,3 & A: $[-0.993,0,-0.1184]$ & A: $(0.06,-0.998 i)$ \\
Fig. S6& B: $[0.993,0,-0.1184]$ & B: $(0.998,-0.06 i)$ \\
\hline Fig. S4 & A: $[-0.16,-0.48,-0.86]$ & A: $(0.648,-0.37-0.665i)$ \\
Fig. S8 & B: $[0.45,0.77,0.46]$ & B: $(0.8515,0.45+0.269i)$ \\
\hline Fig. S5 & A: $[0.43,-0.9,-0.03]$ & A: $(0.846,-0.534-0.018i)$ \\
Fig. S7 & B: $[0.38,0.91,-0.17]$ & B: $(0.83,0.5473-0.1022i)$ \\
\hline
\end{tabular}
\caption{Normalized Jones vectors and Stokes parameters for polarizations of QNM (supported by different structures studied in both the main Letter and this Supplemental Material) radiations opposite to the incident direction.} 
\label{table} 
\end{table}

\begin{figure}[htbp]
\centerline{\includegraphics[width=12cm]{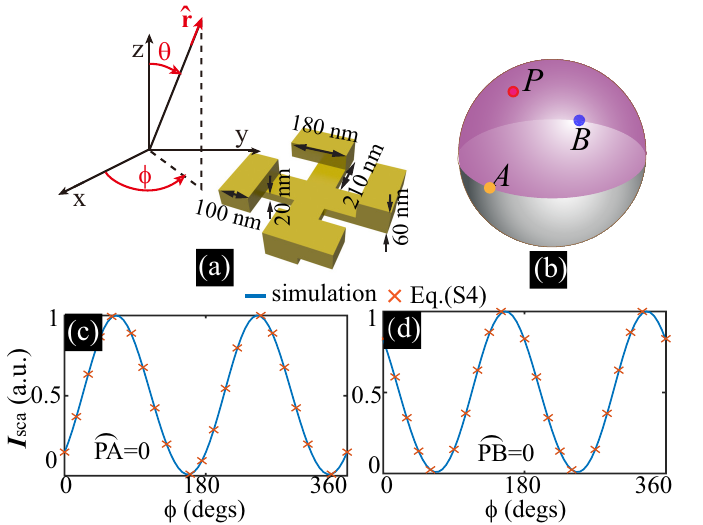}} \caption{\small (a) A gold particle with 4-fold rotation symmetry and a spherical coordinate system parameterized by $\mathbf{r}=(r,~\theta,~\phi)$. It is the same particle as shown in Fig. 2(a) of the main Letter. (b)  Poincar$\mathrm{\acute{e}}$ sphere on which P, A and B represent respectively the incident polarization and polarizations of the mode radiations opposite to the incident direction. Scattering intensity distributions on the $\mathbf{x}$-$\mathbf{y}$ plane: (c) P overlaps with A; (d) P overlaps with B. Refer to Table \ref{table} for specific parameters of A and B.}
\label{fig-selective}
\end{figure}

\section{(\textbf{\uppercase\expandafter{\romannumeral4}}). Selective QNM excitations}
\label{selective-excitation}

In this section, we demonstrate selective mode excitations for the degenerate QNMs supported by the particle shown in Fig. 2(b) of the main letter [re-shown here in Fig.~\ref{fig-selective}(a)].
The particle consists of gold and exhibits four-fold rotation symmetry that secures a pair of degenerate QNMs with central resonant angular frequency $\omega_1 = 8.0645\times10^{14}~\rm{rad}/s$. We shine plane waves along $\mathbf{+z}$  with  $\omega=\omega_1$ and track scattering distributions on the $\mathbf{x}$-$\mathbf{y}$ plane.  The radiations of the two QNMs along -$\mathbf{z}$ (opposite to the incident direction) are almost linearly polarized along $\mathbf{x}$ and $\mathbf{y}$, with the corresponding normalized Stokes parameters~\cite{YARIV_2006__Photonics} being respectively $S_1\approx\pm1$ [Fig.~\ref{fig-selective}(b)]. The normalized scattering intensity distributions on the $\mathbf{x}$-$\mathbf{y}$ plane [parameterized by the azimuthal angle $\phi$ as shown in Fig.~\ref{fig-selective}(a)] are shown in Fig.~\ref{fig-selective}(c) and \ref{fig-selective}(d), for incident linear polarizations along $\mathbf{x}$ ($\arc{PA}\approx0$, $\arc{PB}\approx\pi$ and thus mode B is not excited: $I_{\rm{sca}} \propto I_{\rm{A}}$) and $\mathbf{y}$ ($\arc{PB}\approx0$, $\arc{PA}\approx\pi$ and thus mode A is not excited: $I_{\rm{sca}} \propto I_{\rm{B}}$), respectively. As is clearly shown in Figs.~\ref{fig-selective}(c) and \ref{fig-selective}(d), radiations of modes A and B [$I_{\rm{A, B}}(\phi)$] agree well with the directly simulated intensity distributions without involving QNMs, confirming the accuracy of our model and also that by properly selecting the incident polarizations, modes can be selectively excited as long as A and B do not overlap. 

\begin{figure}[htbp]
\centerline{\includegraphics[width=12cm]{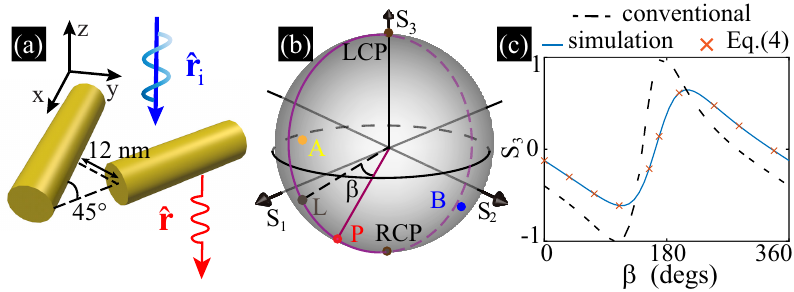}} \caption{\small (a) A pair of non-perpendicular gold cylinders (one twisted by $45^{\circ}$ with respect to the other). (b) Poincar$\mathrm{\acute{e}}$  sphere on which P, A  and B  represent respectively the incident polarization and polarizations of the mode radiations opposite to the incident direction.
(c) Dependence of $S_3$ for scatterings along -\textbf{z} on incident polarizations (perpendicular incidence with $\mathbf{\hat{r}}_{\rm{i}}$ along -\textbf{z}) transversing a great circle parameterized by $\beta$. The incident wavelength $\lambda=0.8~\mu$m. Refer to Table \ref{table} for specific parameters of A and B.}
\label{figure6_SP}
\end{figure}

\begin{figure}[htbp]
\centerline{\includegraphics[width=12cm]{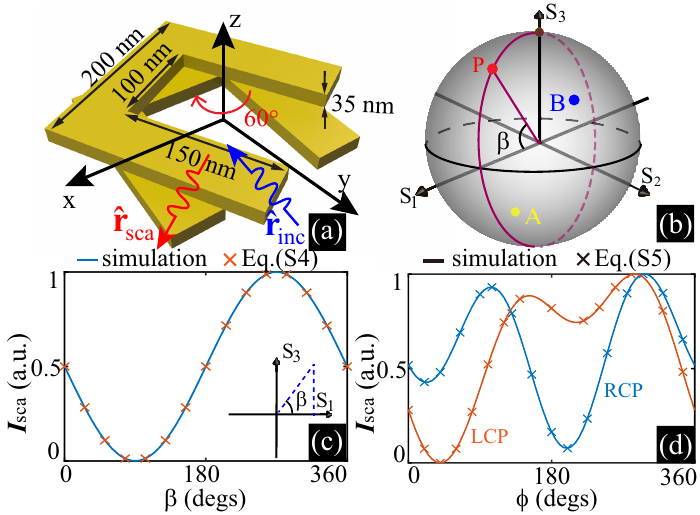}} \caption{\small (a) A pair of gold split-ring resonators, and two directions of interest ($\mathbf{\hat{r}}$ and $\mathbf{\hat{r}}_{\rm{i}}$) are also marked. The corresponding A and B  for this $\mathbf{\hat{r}}_{\rm{i}}$ are shown in (b). The dependence of the directional scattering intensity along $\mathbf{\hat{r}}$ on incident polarizations is shown in (c), with P transversing a great circle on the Poincar$\mathrm{\acute{e}}$ sphere ($S_2=0$), covering LCP ($\beta=\pi/2$) and RCP ($\beta=3\pi/2$). (d) Scattering intensity distributions on the $\mathbf{x}$-$\mathbf{y}$ plane with LCP and RCP incidences along the marked incident direction. The incident wavelength $\lambda=1.66~\mu$m and refer to Table \ref{table} for specific parameters of A and B.}
\label{figbisrr}
\end{figure}

\section{(\textbf{\uppercase\expandafter{\romannumeral5}}). Non-perpendicular bi-cylinder scattering}

We have also studied another bi-cylinder (non-perpendicular with respect to each other) scattering configuration shown in Fig.~\ref{figure6_SP}(a). The QNM radiations along the opposite incident direction (+\textbf{z}) are not linear and their positions are indicated in (b). In Fig.~\ref{figure6_SP}(c) the evolution of the polarization (in terms of $S_3$) for the forward scatterings along -\textbf{z} with varying incident (along -\textbf{z}) polarizations. Here the incident polarization P [see (b)] locates on a great circle parameterized by $\beta$ ($\beta=0$ corresponds to state L that locates on \arc{AB}) and  equally bisects \arc{AB} (\arc{PA}=\arc{PB}). It is clear that results from our model agree perfectly well with the simulation results.

\section{(\textbf{\uppercase\expandafter{\romannumeral6}}). Twisted bi-SRR scattering}
In this section, we study a scattering structure shown in Fig.~\ref{figbisrr}(a), which supports a pair of non-degenerate QNMs with central resonant frequencies close to $\omega_2 = 1.1695\times10^{15}~\rm{rad}/s$, at which the angular frequency of the incident wave is fixed ($\omega=\omega_2$). We have identified a scattering direction $\mathbf{\hat{r}}$ ($\phi=41^\circ$ and $\theta=90^\circ$; on the $\mathbf{x}$-$\mathbf{y}$  plane) along which the radiation polarizations of both QNMs are identical. As has already been elaborated, through tuning the relative exciting amplitude and phase, scattering along this direction can be eliminated, similar to that marked in Fig.2(c) in the main Letter. We have selected an incident direction $\mathbf{\hat{r}}_{\rm{i}}$ ($\phi=0$ and $\theta=10^\circ$) and the corresponding positions of A and B  are shown in Fig.~\ref{figbisrr}(b). The evolution of the directional scattering intensity along $\mathbf{\hat{r}}$ with changing incident polarizations covering a great circle ($S_2=0$) is shown in Fig.~\ref{figbisrr}(c), where the scattering is fully suppressed for LCP incidence ($\beta=\pi/2$). We have also tracked the scattering intensity  distributions on the $\mathbf{x}$-$\mathbf{y}$ plane for LCP and RCP incidences, with the results summarized in Fig.~\ref{figbisrr}(d).

\section{(\textbf{\uppercase\expandafter{\romannumeral7}}). Gain particle scattering}
\begin{figure}[htbp]
\centerline{\includegraphics[width=12cm]{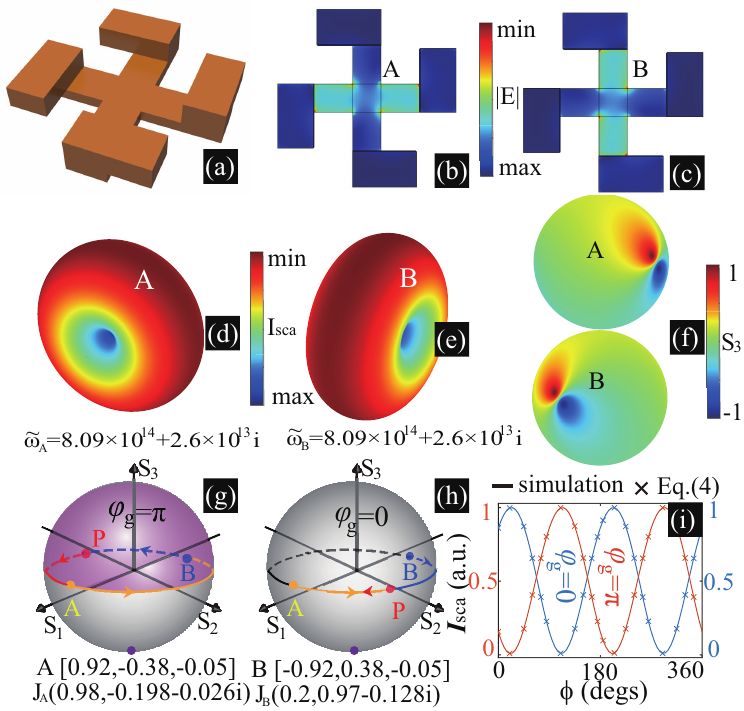}} \caption{\small (a) A gain particle with $4$-fold rotation symmetry that supports a pair of degenerate QNMs A and B. (b) \& (c) Near fields (in terms of total electric field amplitude $\mathbf{|E|}$) of the QNMs. Far-field scattering intensity [(d) \& (e)] and polarization in terms of $S_3$ [(f)] distributions on the momentum sphere.
Scattering intensity distributions on the $\mathbf{x}$-$\mathbf{y}$ plane with P locating at ($0$, $1$, $0$) and ($0$, $-1$, $0$) for (i), which correspond to two orthogonal incident linear polarizations [polarized along $\phi=\pi/4$ ($S_2=1$) and polarized along $\phi=3\pi/4$ ($S_2=-1$)]. The geometric phases are respectively $\varphi_g=0,~\pi$, as shown in (g) \& (h).  In (i) the incident wavelength $\lambda=2.34~\mu$m. In (d) \& (e) the complex eigenfrequencies of the degenerate QNMs are also specified. In (g) \& (h) the normalized Jones vectors and Stokes parameters for polarizations of QNM radiations opposite to the incident direction.}
\label{fig2_SM}
\end{figure}

As has been mentioned already, our theoretical framework relies on the principle of reciprocity only, and thus it is applicable to particles consisting of both passive and gain materials.
In our numerical studies so far,  effective permittivity of gold is fitted from data in Ref.~\cite{Johnson1972_PRB} by the  Drude-Lorentz model: $\varepsilon(\omega)=\varepsilon_{\infty}-\varepsilon_{\infty} \omega_p^2 /\left(\omega^2-\omega_0^2+\right.$ $j \omega \gamma)$, where $\epsilon_{\infty}, \omega_p, \gamma$, and $\omega_0$ denote respectively the permittivity limit of high-frequency, plasma angular frequency, damping coefficient, and resonant angular frequency. Here we study the same particle as shown in Fig. 2 of the main Letter (and also in Figs.~\ref{figure3_SP} and \ref{fig-selective} in this Supplemental Material), and the only difference is that here we replace $\gamma$ by $-\gamma$ in the  Drude-Lorentz formula obtained from experimental data fitting. That is to say, now the particle consists of gain materials [shown in Fig.~\ref{fig2_SM}(a)].  The QNM information (complex eigenfrequencies, near fields, scattered far field intensity and polarization distributions on the momentum sphere) has been summarized in Figs.~\ref{fig2_SM}(b)-(h). Two scattering scenarios (scattering intensity distributions on the $\mathbf{x}$-$\mathbf{y}$ plane) are summarized in Fig.~\ref{fig2_SM}(i), with $\varphi_g=0$ and $\pi$, respectively. It is clear from Fig.~\ref{fig2_SM}(i) that the results obtained with our model agree perfectly well with the simulation results, confirming the validity of theoretical framework for gain particles.

%
%
%
%
%
%
%

\end{document}